\documentclass[a4paper,onecolumn,notitlepage,nofootinbib,twoside,byrevtex,eqsecnum,10pt]{revtex4}

\newcommand{\est}{{\mathcal{E}}}
\newcommand{\hst}{{\mathcal{H}}}
\newcommand{\qst}{{\mathcal{Q}}}

\newcommand{\oms}{\omega_1}
\newcommand{\oma}{\omega_2}
\newcommand{\beq}{\begin{equation}}
\newcommand{\eeq}{\end{equation}}
\newcommand{\bear}{\begin{eqnarray}}
\newcommand{\eear}{\end{eqnarray}}
\newcommand{\degree}{\ensuremath{\,{}^{\circ}}}
\renewcommand{\vec}{\mathbf}

\usepackage{graphicx}
\usepackage{bm}
\usepackage{eurosym}

\begin{document}

\title{
{\underline M}icrowave {\underline A}pparatus for {\underline G}ravitational Waves {\underline O}bservation
}

\author{R. Ballantini}
\affiliation{INFN and Universit\`a degli Studi di Genova, Genova, Italy}
\author{Ph. Bernard}
\affiliation{CERN, Geneva, Switzerland}
\author{S. Calatroni}
\affiliation{CERN, Geneva, Switzerland}
\author{E. Chiaveri}
\affiliation{CERN, Geneva, Switzerland}
\author{A. Chincarini}
\affiliation{INFN and Universit\`a degli Studi di Genova, Genova, Italy}
\author{R.P. Croce}
\affiliation{INFN, Napoli, and Universit\`a degli Studi del Sannio, Benevento, Italy}
\author{S. Cuneo}
\affiliation{INFN and Universit\`a degli Studi di Genova, Genova, Italy}
\author{V. Galdi}
\affiliation{INFN, Napoli, and Universit\`a degli Studi del Sannio, Benevento, Italy}
\author{G. Gemme}
\email{gianluca.gemme@ge.infn.it}
\affiliation{INFN and Universit\`a degli Studi di Genova, Genova, Italy}
\author{R. Losito}
\affiliation{CERN, Geneva, Switzerland}
\author{R. Parodi}
\affiliation{INFN and Universit\`a degli Studi di Genova, Genova, Italy}
\author{E. Picasso}
\affiliation{INFN and Scuola Normale Superiore, Pisa, Italy}
\affiliation{CERN, Geneva, Switzerland}
\author{V. Pierro}
\affiliation{INFN, Napoli, and Universit\`a degli Studi del Sannio, Benevento, Italy}
\author{I.M. Pinto}
\affiliation{INFN, Napoli, and Universit\`a degli Studi del Sannio, Benevento, Italy}
\author{A. Podest\`a}
\affiliation{INFN and Universit\`a degli Studi di Genova, Genova, Italy}
\author{R. Vaccarone}
\affiliation{INFN and Universit\`a degli Studi di Genova, Genova, Italy}

\begin{abstract}
In this report the theoretical and experimental activities for the development of superconducting microwave cavities for the detection of gravitational waves are presented.
\end{abstract}

\maketitle

\tableofcontents

\section*{\label{chap:intro} Introduction}
Existing proposals for next--generation Earth--based gravitational
wave detectors, including advanced interferometers (IFOs) and wideband (nested)
acoustic detectors, are aimed at constructing large--scale
prototypes, with strain sensitivity goals of the order of
$10^{-23}$ in a wide (a few kHz) band. The successful achievement
of these goals will depend on substantial advances to be made,
well beyond the limits of presently achieved figures of merit, in
several critical areas, including (ultra)cryogenics (mK cryostats
with adequate refrigerating power), material science (achieving
extremely high mechanical quality factors at cryogenic
temperatures) and technology (annealing/sintherizing high--quality
deca--ton size objects), electronics (SQL amplifiers), lasers and
optics (high--finesse, large--waist, short optical cavities
performing as displacement readouts at sensitivity levels $\sim
10^{-23}$~m Hz$^{-1/2}$). All these goals are two order of
magnitude beyond the best figures available today, on average.

The MAGO proposal stems from a different perspective, and
emphasizes a relatively new concept. The MAGO design aims at
constructing a relatively large ($N \agt 10$) number of
detectors, relying only on \emph{available and already proven}
technology.

Experience gained on downsized MAGO prototypes suggests that MAGOs
will be relatively cheap, compact--sized, and structurally simple.
Indeed, within the limits of available and already proven
technologies, the proposed MAGO design could allow to construct
compact detectors, featuring strain sensitivities comparable to
those of present day cryogenic acoustical detectors, at a fraction
of their size and cost. This would permit to place several MAGOs
even in a small area (which would be required, e.g., for observing
a stochastic GW background), either co--tuned or frequency
staggered, thanks to the unique bandwidth and center frequency
tuning ease offered by the design.

The potential advantages of many--detector arrays and networks
have been repeatedly emphasized. Similar to radioastronomy,
gravitational wave astronomy will be eventually made possible by
the availability of large assemblies of co--tuned (array) or
frequency staggered (xylophone) detectors featuring robust, high
duty--cycle operation, as a result of constructive simplicity, and
non critical core technologies. The MAGO project and design is
definitely aimed at moving a first, perhaps tiny step toward this
direction.

The following specific technical points are worth being
emphasized:
\begin{itemize}
\item{The individual MAGO bandwidth
and center frequency of operation can be easily tuned well beyond
the operational limits of present day detectors, imposed by laser
shot noise (IFOs) and  Brownian (acoustic) noise;}

\item{The very large parametric--conversion gain which provides the
basic principle of operation of the detector brings the equivalent
temperature of the front-end electronics down to the standard
quantum limit already using  standard, cheap and reliable
off-the-shelf HEMT amplifier ($T \sim 2$ K at $1$~GHz), without
the need of resorting to more sophisticated devices;}

\item{The required electrical quality factor for  the Nb-coated
superconducting cavities, has been safely and routinely achieved
in accelerator technology, and could be realistically improved by
one order of magnitude relatively soon;}

\item{The required mechanical quality factor of the cavities is also
conservative and could be improved;}

\item{The phase noise of the microwave pumping signal is well within
the limits of currently available technology.}
\end{itemize}

The main technology challenge is thus essentially related to the
design and testing of a cryostat capable of removing the heat
produced in the cavity walls by the microwave pumping signal (a
few Watts, typically) so as to maintain an operating temperature $\sim
2$~K, while guaranteeing adequate acoustical isolation and very
low intrinsic noise.

On the other hand, it should be noted that the MAGO design has
{\em no} a--priori intrinsic limitation which could prevent it
from reaching performance figures, both in terms of strain
sensitivity and bandwidth, comparable to those of proposed next
generation large--scale detectors (including advanced IFOs and
wideband nested acoustic detectors), provided one pushes the
relevant critical design figures (cryogenics, mass, mechanical
quality factor, readout noise) up to comparable levels.

In particular, if we were able to design and operate the cryostat
down to the mK range, the MAGO strain sensitivity could easily
reach the $h \sim 10^{-23}$ level, while preserving MAGO's almost
unique features in terms of center frequency and bandwidth easy
tunability. As of today, the best performing 100mK cryostats are
capable of delivering only a few hundred mW. We note that next
generation optically read--out wideband nested--acoustical
detector would face  the same technological challenge, among
others, in order to reach their foreseen sensitivity goal.

\section{\label{chap:overv}Experiment Overview}
In the last decades, several laboratories all around the world
have promoted an intense effort devoted to the direct detection of
gravitational waves. The detectors, both those in operation and
those being developed, belong to two conceptually different
families, massive elastic solids (cylinders or spheres)
\cite{websbarre} and Michelson interferometers \cite{webinterf}.
Both types of detectors are based on the mechanical coupling
between the gravitational wave and a test mass, and in both types
the electromagnetic field is used as motion transducer.

In a series of papers, since 1978, it has been studied how the
energy transfer induced by the gravitational wave between two
levels of an electromagnetic resonator, whose frequencies $\oms$
and $\oma$ are both  much larger than the characteristic angular
frequency $\Omega$ of the g.w., could be used to detect
gravitational waves \cite{bppr1,bppr2}. The energy transfer is
maximum when the resonance condition $|\oma - \oms| = \Omega$ is
satisfied. This is an example of a frequency converter, i.e. a
nonlinear device in which energy is transferred from a reference
frequency to a different frequency by  an external pump signal.

In the scheme suggested by Bernard et al. the two levels are
obtained by coupling two identical high frequency cavities
\cite{bppr2}. Each resonant mode of the individual cavity is then
split in two modes of the coupled resonator with different spatial
field distribution. In the following we shall call them the {\em
symmetric} and the {\em antisymmetric} mode. The frequency
difference of the two modes (the detection frequency) is
determined by the coupling, and can be tuned by a careful
resonator design. An important feature of this device is that the
detection frequency does not depend on its mechanical properties
(dimensions, weight and mechanical modes resonant frequencies),
though, of course, the detector {\em can} be tuned so that the
mode splitting equals the frequency of a mechanical resonant mode.
The sensitivity in this and other experimental situations will be
discussed in the following. Since the detector sensitivity is
proportional to the electromagnetic quality factor, $\qst$, of the
resonator, superconducting cavities should be used for maximum
sensitivity.

\begin{figure}[hbt]
 \begin{center}
\includegraphics[scale=0.4]{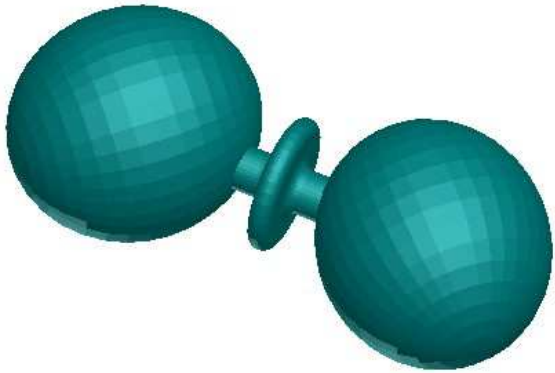}
 \caption{
 \label{fig:paco2}
 Artistic view of the coupled spherical cavities with the central tuning cell}
 \end{center}
 \end{figure}

An R\&D effort, started in 1998, was completed at the end of 2003
 \cite{rsi, cqg}. Its main
objective was the development of a {\em tunable} detector of small
harmonic displacements based on two coupled superconducting
cavities. Several cavity prototypes (both in copper and in
niobium) were built and tested, and finally a design based on two
spherical cells was chosen and realized (Fig. \ref{fig:paco2}).
The detection frequency, i.e. the frequency {\em difference}
between the symmetric and antisymmetric modes, was chosen to be:
$\oma -\oms \approx 10$ kHz (the frequency of the modes being
$\omega_{1,2} \approx 2$ GHz). An electromagnetic quality factor
$\qst > 10^{10}$ was measured on a prototype with fixed coupling.

\begin{figure}[hbt]
 \begin{center}
\includegraphics[scale=0.4]{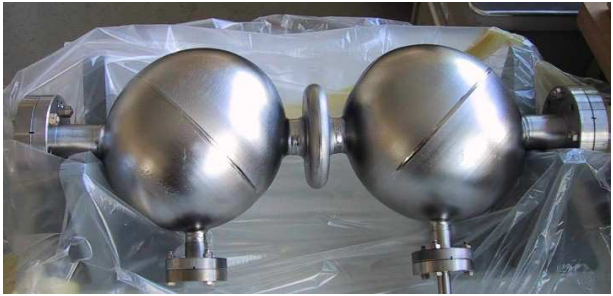}
 \caption{
 \label{fig:tunable}
 Niobium spherical cavities (variable coupling)
 }
 \end{center}
 \end{figure}

The tuning system was also carefully studied. The coupling
strength, and thus the tuning range, is determined by the diameter
of the coupling tube and by the distance between the two sperical
cells. A central elliptical cell, which can easily be streched and
squeezed, was found to provide a tuning range of several kHz
(4--20 kHz in the final design). A prototype with the central
elliptical cell was built and tested (Fig.
\ref{fig:tunable}).

The system was also mechanically characterized, and the mechanical
resonant modes in the frequency range of interest were identified.
In particular the quadrupolar mode of the sphere was found to be
at 4 kHz, in good agreement with finite elements calculations.

The detection electronics was designed. Its main task is to
provide the rejection of the symmetric mode component at the
detection frequency. A rejection better than 150 dB was obtained
in the final system.

Starting from the results obtained in the last six years, we are
now planning to design and realize an experiment for the detection
of gravitational waves in the 4--10 kHz frequency range. Our main
task is the design and construction of the refrigerator and of the
cryostat (including the suspension system), which houses the
coupled cavities. The refrigerator must provide the cryogenic
power needed to keep the superconductiong cavities at $T \approx
1.8$ K (approx. 10 Watts) without introducing an excess noise from
the external environment. A design based on the use of subcooled
superfluid helium is being invesigated.

In the following a detailed description of the various issues
aforementioned will be given. Expected system sensitivity will
also be discussed.

\section{\label{chap:motiv}Physics Motivation}
The spectrum of gravitational waves of cosmic origin targeted by
currently operating or planned detectors spans roughly \footnote{
%+++++++++++++++++++++++
We leave out deliberately the ELF ($10^{-18}-10^{-15}$ Hz)
radiation resulting from inflation--enhanced primordial
gravitational fluctuations, expected to show up in the
polarization anisotropy of the cosmic microwave background (M.
Kamionkowski, A. Jaffe, Int, J. Mod. Phys. A16, 116, 2001),  the
VLF ($10^{-7}-10^{-9}$ Hz) radiation possibly resulting from
extremely massive black-hole systems and early--universe processes
(A.N. Lommen and D.C. Backer, Astrophys. J., 562, 297, 2001,
astro-ph/0107470), exotic electromagnetic--to--gravitational wave
conversion mechanisms, which might originate gravitational
radiation in the VHF to SHF bands (Fang-Yu Li and Meng-Xi Tang,
Int. J. Mod. Phys D11, 1049, 2002) and the relic gravitational
radiation (B. Allen and R. Brustein, Phys. Rev. D55,  3260,
1997).}
%+++++++++++++++++++++++
from $10^{-4}$ to $10^4$ Hz.

The $f \leq 10^{-1}$ Hz region of the GW spectrum, including
galactic binaries \cite{gal_bin}, (super)massive BH binary
inspirals and mergers \cite{SMBH}, compact object inspirals and
captures by massive BHs \cite{capture}, will be thoroughly
explored by LISA \cite{LISA},  which might be hopefully flown by
year 2015. Ground based interferometers and acoustic detectors
(bars and spheres) will likewise co--operate in exploring the $f
\geq 10^1 Hz$ region of the spectrum, including compact binary
inspirals and mergers \cite{CB_inspiral}, supernovae and newborn
black-hole ringings \cite{SN_BHring}, fast-spinning
non-axisymmetric neutron stars \cite{NS}, and stochastic GW
background \cite{stochastic}.

The whole spectral range from $10^{-4}-10^4$ Hz, however, is far
from being covered with uniform sensitivity. Plans are being made for
small--scale LISA--like space experiments (e.g., DECIGO,
\cite{DECIGO}) aimed at covering the frequency gap $10^{-1} -10^1$
Hz  between LISA and terrestrial detectors.

Several cryogenic/ultracryogenic acoustic (bar) detectors are also
operational, including ALLEGRO \cite{ALLEGRO}, AURIGA
\cite{AURIGA}, EXPLORER \cite{EXPLORER}, NAUTILUS \cite{NAUTILUS},
and NIOBE \cite{NIOBE}. They are tuned at $\sim 10^3$ Hz, with
bandwidths of a few tens of Hz, and minimal noise PSDs of the
order of $10^{-21}$  Hz$^{-1/2}$.

Intrinsic factors exist which limit the performance of both IFOs
and acoustic detectors in the upper frequency decade ($f \agt
10^3$ Hz) of the spectrum.

The high frequency performance of laser interferometers is limited
by the $\propto f^2$ raise of the laser shot-noise floor. While it
is possible to operate IFOs in a resonant (dual) light-recycled
mode, for narrow-band increased-sensitivity operation, the pitch  frequency should be kept below the
suspension violin-modes \cite{violin}, typically clustering near
and above $\sim 5 \cdot 10^2$ Hz.

Increasing the resonant frequency of acoustic detectors (bars,
spheres and TIGAs), on the other hand, requires decreasing their
mass $M$. The high frequency performance of bars and spheres is
accordingly limited by the  $\propto M^{-1/2}$   dependance of the
acoustic detectors' noise PSD.

The next generation of resonant detectors will be probably spheres
or TIGAs  (Truncated Icosohedral Gravitational Antennas,
\cite{Merk_John}) \footnote{
%+++++++++++++++
Spheres and TIGAs share the nice feature of being inherently
omnidirectional, and should allow to reconstruct the direction of
arrival and polarization state of any detected gravitational wave,
by suitably combining the outputs of  transducers gauging the
amplitudes of the five degenerate quadrupole sphere modes (C. Zhou
and P.F. Michelson, Phys. Rev. D51, 2517, 1995).}
%+++++++++++++++
The MINIGRAIL \cite{MINIGRAIL} spherical prototype\footnote{
%+++++++++++++++
A cryogenic solid-CuAl sphere resonating at $\sim 10^3$ Hz would
be $\sim 4$ m $\emptyset$ and weigh in excess of $100$ tons. The
related technological challenges could be alleviated by resorting
to hollow geometries (J.A. Lobo, Class. Quantum Grav., 19, 2029,
2002).}
%+++++++++++++++
experiment  under construction at Leiden University (NL), as well
as its twins planned  by the Rome group \cite{SFERA}  and at
S\~{a}o Paulo, Brazil \cite{Mario_Schenberg},  is a relatively
small (CuAl (6\%) alloy, $\emptyset$ 65cm, 1.15ton) spherical
ultracryogenic (20mK) detector with a 230Hz bandwidth centered at
3250Hz, and a (quantum limited) strain sensitivity of  $h \sim
4\cdot 10^{-21}$. Spherical (or TIGA) detector might achieve
comparable sensitivities up to $f \sim 4 \cdot 10^3$ Hz.

Summing up, the GW spectrum  below $f  \sim 10^3$ Hz might be
adequately covered by ground-based and space-borne
interferometers. The range between $10^3$ Hz and $\sim 4\cdot
10^3$ Hz could be sparsely covered by new-generation acoustic
detectors. The high frequency part ($f \agt 4 \cdot 10^3$ Hz) of
the gravitational wave spectrum of cosmic origin is as yet
completely uncovered. Within this band, GW sources might well
exist and be observed \footnote{
%++++++++++++++++++++++++
A well known back-of-an envelope estimate (motion at the speed of
light along body-horizon circumference) gives the following upper
limit for the spectral content of  gravitational waves originated
in a process involving an accelerated mass $\sim M$: $f_{sup}
\alt \frac{c^3}{4\pi G M} \sim 10^4 (M_{\odot}/M)$ [Hz] .}.
%++++++++++++++++++++++++
Indeed, the ultimate goal of gravitational-wave astronomy is the
discovery of {\em new} physics.  In this spirit, the very
existence of gravitational wave sources of as yet unknown kind
could not be excluded a-priori.

The above brings strong conceptual and practical motivations for
the MAGO proposal. The MAGO design is easily scalable, and may be
constructed to work at {\em any} chosen frequency in the range
$10^3-10^4$ Hz, with uniform (narrowband)  performance. On the
other hand, the MAGO instrument appears to be comparatively cheap
and lightweigth, thus allowing to build as many  detectors as
needed to ensure adequate covering of the high frequency ($f \agt
4\cdot 10^3$ Hz)  GW spectrum. In view of their limited cost,
MAGOs might also be nice candidates for many--detector networks,
to achieve very low false alarm probabilities in coincidence
operation. In addition MAGO--like detectors operating at $f \sim
10^3$ Hz might hopefully provide coincident observations with both
acoustic detectors and IFOs, based on a {\em different} working
principle.

Before all this might come into reality, it will be necessary to
build and operate one or more MAGO prototypes so that some basic
issues might be efficiently addressed and solved, viz.:
\begin{itemize}
\item{efficient decoupling from platform $\rightarrow$ suspensions design;}
\item{efficient and quiet cooling to 1.8K $\rightarrow$ cryostat design;}
\item{efficient readout  $\rightarrow$ microwave feeding and tapping networks, and low noise amplifier design.}
\end{itemize}
In parallel, a start--to--end simulation codes should be
implemented, in order to tune all design parameters  for best
operation. In particular, criteria for obtaining the best tradeoff
between detector bandwidth and noise levels should be
investigated, with specific reference to selected classes of
sought signals.

\section{Sources in the range $10^3$~Hz--$10^4$~Hz}
\subsection{\label{sec:mergers} Mergers}
A well known upper frequency limit for the spectrum of
gravitational radiation originated in a process involving an
accelerated mass $M$ is given by \beq f_{sup}\sim\frac{c^3}{4\pi
GM}\sim10^4\frac{M_{\odot}}{M}~\mathrm{[Hz]} \eeq corresponding to
the rather extreme assumption of  motion at the speed-of-light at
the gravitational body horizon. These conditions are met almost
{\em verbatim} in compact binary mergers, which have been
accordingly indicated as the most reliable sources of
gravitational waves in the frequency range from $10^3$~Hz to
$10^4$~Hz \cite{merger_hughes}. Gravitational waves from mergers
are expected to carry rich astrophysical information on the nature
of the coalescing stars, and in particular, as to whether exotic
(e.g. quark) matter is involved \cite{merger_strange}.
%The corresponding event--rates are encouraging \cite{merger_rates},
%being of the order of tens of events per year throughout the
%galactic halo.

In view of our still relatively limited knowledge of the relevant
waveforms \cite{merger_waveform}, it is difficult to estimate the
related signal to noise ratios. However mergers will follow an
inspiral phase, for which more reliable estimates can be made.

To learn as much as possible from the waves generated during the
merger, broadband GW detectors should be supplemented by {\em
narrow band} detectors. A "xylophone" of narrow band detectors
would probe features of the merger waveform which should be robust
in the sense that they would not require detailed modeling of the
waves' phasing. To make best use of these detectors, the network
should be designed in an optimal way: the narrowband detectors
should be tuned, in concert with the broadband detectors, so that
the network of all detectors is most likely to provide new
information about merger waves \cite{merger_hughes}. Practical
considerations such as cost, available facility space and
tunability strongly favor the use of a network of MAGOs.

\subsubsection{Event rates}
Crude estimates of NS--NS merger event-rates in our Galaxy are
obtained by dividing the estimated number of NS--binaries in our
Galaxy by the average time they take to merge. Multiplying this
figure by the number of galaxies (or fraction of Galaxy disk
volume) within the reach (visibility distance) of a given detector
gives a crude estimate of the {\em detectable} event-rate.

This exercise has been repeated through the last decade by several
Authors \cite{merger_rate0}--\cite{merger_rateN}.

The most refined analysis available to date, including the
simulation of selection effects inherent in all relevant radio
pulsar surveys, and a Bayesian statistical analysis for the
probability distribution of the relevant merger rate, has been
presented by Kim, Kalogera and Lorimer \cite{merger_rateN}. They
estimate the NS-NS merger event-rate in the Galaxy between
$0.2\times 10^{-5}$~Mpc$^{-3}$~yr$^{-1}$ and $6.0\times
10^{-5}$~Mpc$^{-3}$~yr$^{-1}$ depending on the assumed pulsar
population model, and provide a NS-NS merger event-rate
probability distribution yielding a most likely value for the
NS-NS merger rate between $0.3\times 10^{-5}$~Mpc$^{-3}$~yr$^{-1}$
and $1.7 \times 10^{-5}$~Mpc$^{-3}$~yr$^{-1}$ events per year at a
$68\%$ confidence level, in agreement with previously obtained
orders of magnitude.

\begin{figure}[hbt]
\begin{center}
\includegraphics[scale=0.5]{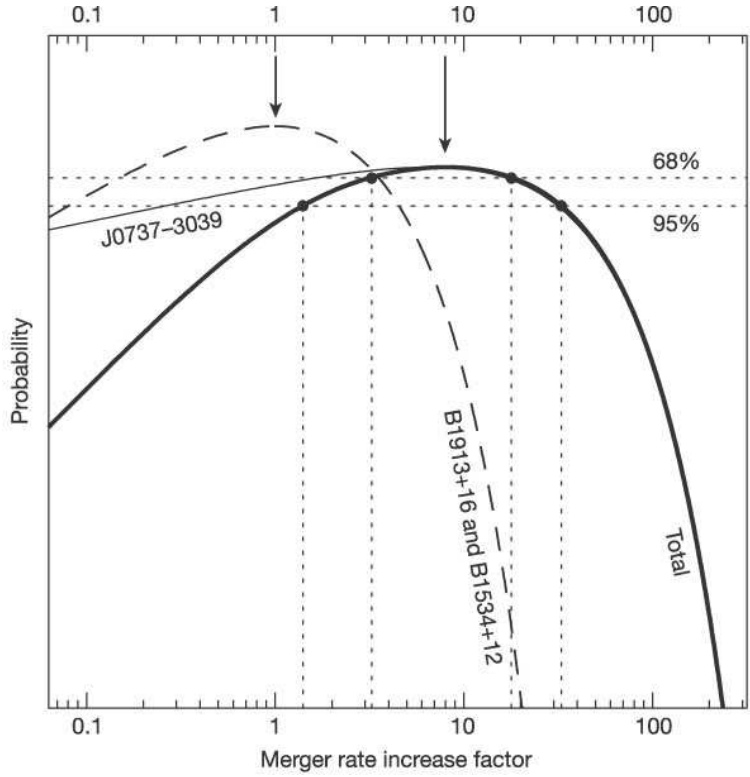}
\caption{Probability density function (pdf) for the increase in
the double--neutron--star merger rate resulting from the discovery
of PSR J0737--3039. The dashed line represents the pdf
corresponding to the \emph{old} merger rate (before the discovery
of PSR J0737--3039). The heavy solid line represents the pdf
corresponding to the \emph{new} merger rate (from
\cite{new_NSNS}). \label{fig:A} }
\end{center}
\end{figure}

Interest in mergers on behalf of the GW Detectors' Community has
been revived  by the recent (published dec. 2003) discovery by
Burgay and co-workers of a new, highly relativistic NS-NS binary
(PSR J0737--3039) with an orbital period of $2.4$~hr and an
estimated lifetime of only $85$~Myr \cite{new_NSNS}. The estimated
NS-NS merger event-rates depend critically on the {\em shortest}
observed NS-binary lifetime. This observation alone suffices to
boost the estimated NS-NS merger event rate in the Galaxy by a
factor between $4$ and $7$ at the $68\%$ confidence level, as
shown in Fig. \ref{fig:A} \cite{new_NSNS}.

The remarkable weakness of PSR J0737-3039 radio signals, despite
its relative nearness ($\sim 600$~pc) to Earth, further suggests
that there might be a plethora of {\em as yet undetected}
short-lived NS-NS binaries in the Galaxy. This is the dominant
source of uncertainty in estimating NS-NS merger rates, which
might accordingly be $\sim 10^2$ larger than estimated, as
stressed in \cite{merger_rate4}, where it is argued that all other
modeling uncertainties build up only to an uncertainty factor of
the order of unity.

In order to translate the above event rates into {\em observable}
event rates, one has to estimate the visibility distance of the
MAGO detector. We define here the visibility distance as the
distance of a source yielding SNR$=1$.

To determine the visibility distance of MAGO (and/or a MAGO array,
or xilophone), one may adopt the std. (conservative) procedure
appropriate for the case where detailed knowledge of the waveform
is not available. One accordingly has to compare the squared
sought signal {\em characteristic amplitude},
$$
h^2_c=\frac{2}{\pi^2 r^2} \frac{dE}{df}
$$
to the squared detector's {\em r.m.s. noise ampltude},
$$
h^2_n=\frac{fS(f)}{\langle F^2 \rangle}
$$
so as to recast the signal to noise ratio into the form
$$
\mathrm{SNR}=\left[\int d(\log f) \frac{h^2_c(f)}{h^2_n(f)}
\right]^{1/2}
$$
where $S(f)$ is the detector noise power spectral density (PSD),
$F$ is the detector's directivity function, $\langle \cdot
\rangle$ denotes averaging over all directions, and $dE/df$
describes the spectral energy content of the sought signal. Based
on reasonable assumptions about the transition from inspiral to
merger, we might use for this latter the formula (in geometrized
units)
$$
\frac{dE}{df}\sim \displaystyle{ \left\{
\begin{array}{l}
0.91 M^2 \times \left(4\mu/M\right)^2,~~~f \in (f_{merge},f_{ring})\\
0,~~~\mathrm{elsewhere}
\end{array}
\right. }
$$
first proposed in \cite{Flanagan}, where $f_{merge}\sim
f^{(GW)}_{LSCO}\sim 4\times 10^3 (M_\odot/M)$~Hz and $f_{ring}\sim
2.6\times 10^4 (M_\odot/M)$~Hz is the quasi-normal mode ring-down
frequency of the final collapsed object \cite{Flanagan}.

\begin{figure}[hbt]
\begin{center}
\includegraphics[scale=0.75]{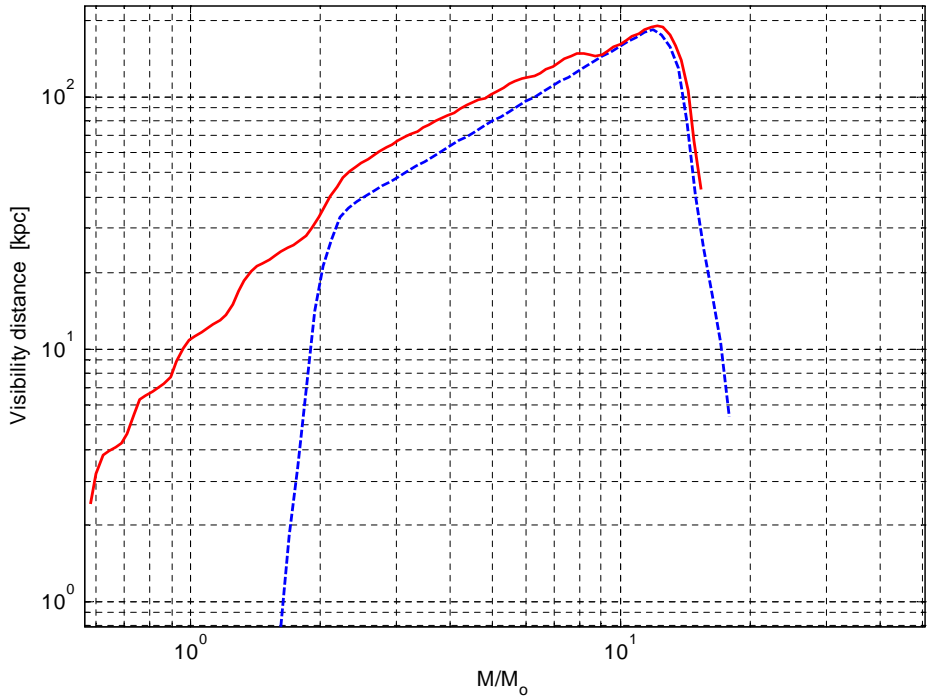}
\caption{Visibility distance of one MAGO (dashed line) and of a
\emph{xylophone} of six MAGOs covering the 2--8 kHz frequency
range (solid line) vs. total mass of the merger (in solar mass
units). \label{fig:merger} }
\end{center}
\end{figure}

For a medium-size MAGO ($S_h^{1/2}\sim 6\times
10^{-21}$~Hz$^{-1/2}$ @ $2$~KHz in a $350$~Hz bandwidth) the
corresponding visibility distance (SNR$=1$) will be $\sim 0.2$~Mpc
for a $M\sim 10 M_\odot$ merger with a sharp cut--off at $M\sim 2
M_\odot$. With a MAGO xylophone covering the 2--8~KHz band at the
same PSD level, the mass range could be extended down to $M\sim
0.6 M_\odot$. (see fig. \ref{fig:merger}).

\subsection{\label{sec:bursts}Bursts}
\begin{figure}[hbt]
\begin{center}
\includegraphics[scale=0.75]{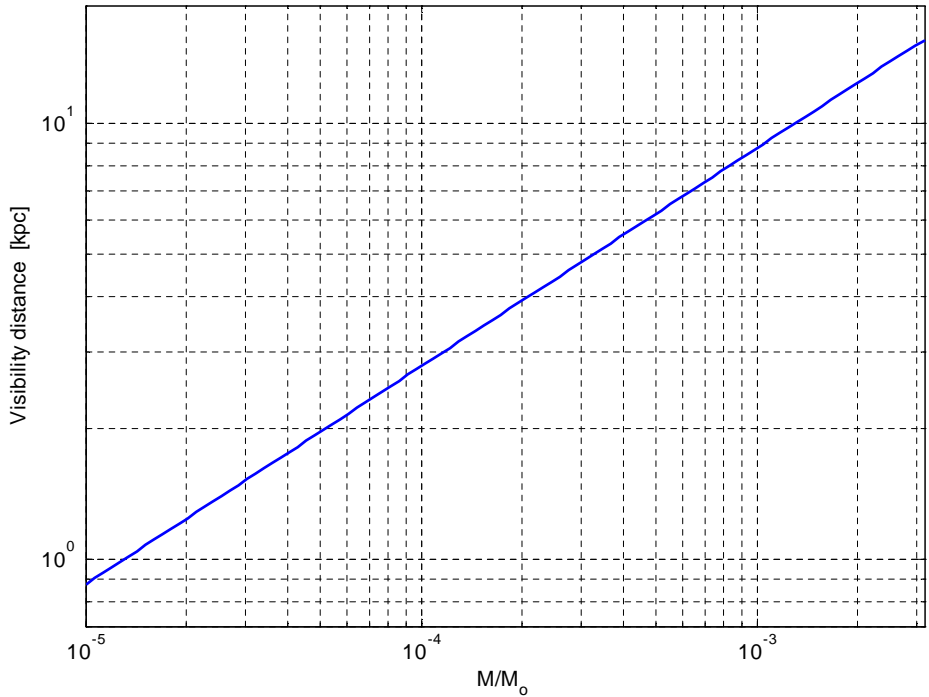}
\caption{Visibility distance of one MAGO vs. burst energy (in
solar mass units). \label{fig:burst} }
\end{center}
\end{figure}

The characteristic noise amplitude of one MAGO is $h_n(f_c)\sim
7\times 10^{-19}$ at the characteristic frequency $f_c \sim
2$~kHz. This accounts for the isotropic conversion into GW energy
of $8\times 10^{-4}$ solar masses located at a distance of 8~kpc
(SNR=1) (see fig. \ref{fig:burst}).

Recently \cite{pizzella} a coincidence excess was found among the
data of the resonant bars EXPLORER and NAUTILUS, when the
detectors are favorably oriented with respect to the galactic
disk. The observed coincidence corresponds to a conventional burst
with amplitude $h_c \sim 2\times10^{-18}$ and to the isotropic
conversion into GW of $4\times10^{-3} M_\odot$, with sources
located in the Galactic Center. Still unknown phenomena other than
GWs, though, cannot be ruled out as causes of the observed events
\cite{pizzella}. A possible way to distinguish GWs from other
sources is the use of several, different, detectors with good
sensitivity. From this point of view the use of a network of MAGOs
could contribute in clarifying this subject.

\subsection{\label{sec:macho}MACHO Binaries}
\label{sec:MACHO} The gravitational wave frequency corresponding
to the last stable circular orbit, loosely marking the transition
from inspiral to plunge in binary coalescence, is roughly
\cite{Finn_vis}: \beq f^{(GW)}_{LSCO} \sim 3976
\frac{M_{\odot}}{M}~[\mathrm{Hz}] \eeq for a (symmetric, non
spinning, circular orbit) binary with total mass $M$.

The spectral window of MAGOs would thus be highly appropriate to
observe gravitational waves from  black-hole MACHO  binaries
\cite{MACHOs} with a typical total mass of $0.6 M_\odot\alt
M\alt 2 M_\odot$. It is speculated \cite{MACHOs2} that one half
of the galactic halo mass (corresponding to $\sim 3\times 10^{11}
M_\odot$) consists of MACHOs with a typical mass of $M\sim
0.3M_{\odot}$. Among these, $\sim 30\%$ should be bound in binary
systems, formed at the same time ($10^{10}$~yr) the galaxy was
born  and uniformly distributed across the halo. The rate of MACHO
binary coalescence could be of the order of $10$ events per year
\cite{acu_Finn} in the galactic halo.

\begin{figure}[hbt]
\begin{center}
\includegraphics[scale=0.5]{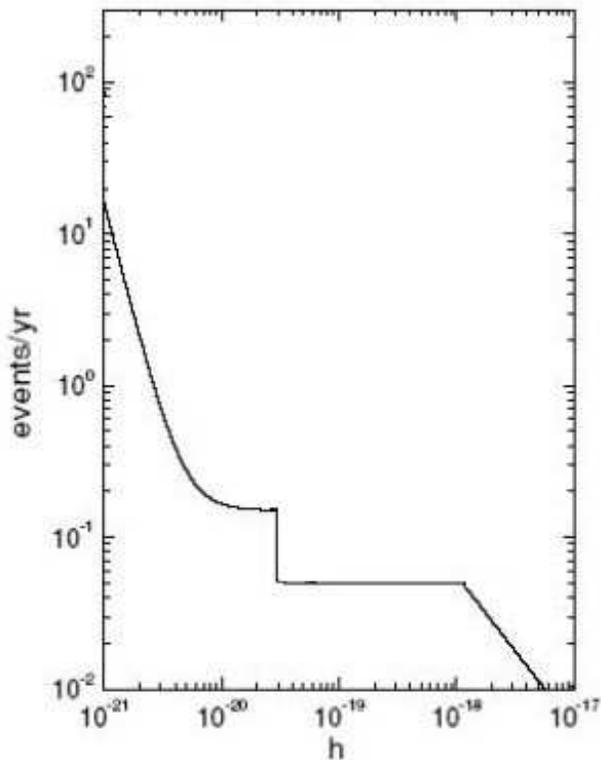}
\caption{BH--MACHO binary inspiral (and merger) observable event
rates for the (advanced) Brazilian spherical antenna "Mario
Schenberg" as a function of the burst sensitivity (from
\cite{MACHO_Brazil}). \label{fig:B} }
\end{center}
\end{figure}

A nice representation of current expectations for BH--MACHO binary
inspiral (and merger) observable event rates has been recently
given by J.C.N. de Araujo and co--workers \cite{MACHO_Brazil} for
the (advanced) Brazilian spherical antenna "Mario Schenberg" as a
function of the burst sensitivity, and is shown in fig.
\ref{fig:B}. The burst sensitivities $h$ in fig. \ref{fig:B} are
readily translated into visibility distances $d$ by noting that
$h(d_1)/h(d_2)=d_2/d_1$.

For visibility distances up to the Galaxy's  border ($d \sim
20$~Kpc) the event rate increases linearly up to $\sim
0.05$~yr$^{-1}$. This accounts for the linear segments on the
right side of fig. \ref{fig:B}. From the border of the Galaxy up
to the outskirts of $M31$ and $M32$, at $d\sim 700$~Kpc,
intergalactic BH--MACHOs might be visible. Their contribution,
however, would be negligible under the assumption that they follow
the distribution of dark matter. This corresponds to the plateau
in figure \ref{fig:B}. As the visibility distance overpasses $M31$
and $M32$, the event--rate contribution of these latter is added,
corresponding  to the step in fig. \ref{fig:B}. For larger
distances, beyond $1$~Mpc, the background BH--MACHOs' contribution
to the event rate becomes  eventually  dominant. The relevant
event--rate is  obtained using the coalescency rate computed in
\cite{BHMACHO_Nakamura} (leftmost region in fig. \ref{fig:B}).

The visibility distance (SNR=1) for observing a coalescing
$0.5M_{\odot}+0.5M_{\odot}$ MACHO binary with a MAGO
($S_h^{1/2}=6\times 10^{-21}$ between $1850$~Hz and $2150$~Hz)
would be $\sim 18$~kpc. The maximum visibility distance for this
detector is 25 kpc for observing a $0.9M_{\odot}+0.9M_{\odot}$
MACHO binary\footnote{The visibility distance of the initial
"Mario Schenberg"  (aka mini--GRAIL, SFERA) antenna
($S_h^{1/2}\sim 2\times 10^{-21}$~Hz$^{-1/2}$ @ $3.2$~kHz in a
$50$~Hz bandwidth) will be $d \sim 17$~kpc at SNR=1 for a
$0.5M_\odot +0.5 M_\odot$ BH--MACHO.}.

\begin{figure}[hbt]
\begin{center}
\includegraphics[scale=2.5]{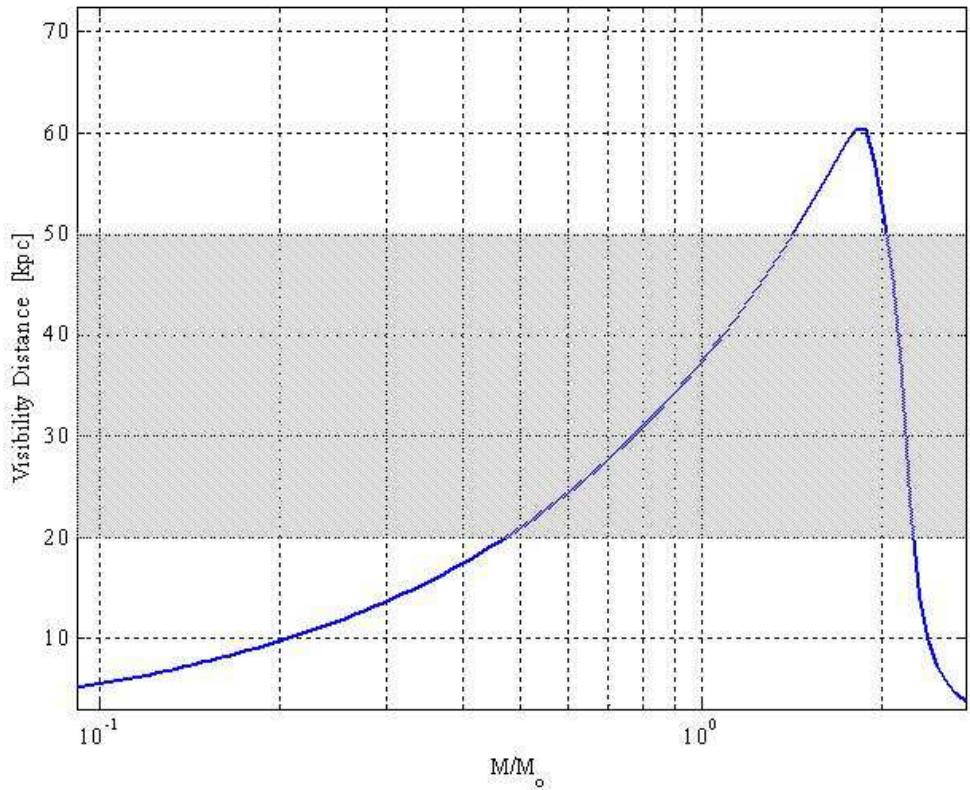}
\caption{Visibility distance of an array of six identical MAGOs
vs. the binary total mass. The typical distance of the galactic
halo is also shown (gray area). \label{fig:MACHO6} }
\end{center}
\end{figure}

Furthermore, in view of MAGO's relatively simple, lightweight and
hopefully cheap features, one might think of building arrays of
many MAGOs, either identical or differently tuned detectors ({\em
xylophone}) whose response would allow to deduce the relevant
source parameters (chirp-mass etc.). Fig. \ref{fig:MACHO6} shows
the visibility distance of both six {\em identical} MAGOs as a
function of the binary total mass.

\subsection{\label{sec:pulsar}Submillisecond Pulsars}
The existence of sub-millisecond pulsars has been speculated by
various Authors. It is well known that the (lower) limiting spin
period of a neutron star depends on the assumed equation of state,
being $\sim 600 \,\mu\mathrm{s}$ for the softest one
\cite{breakup}. Submillisecond pulsars, if any, could be strange
(quark) stars \cite{quark} as well. Known millisecond pulsars are
believed to be "recycled" neutron stars spun-up by accretion from
the companion, in low-mass X-ray binary systems (LMXB)
\cite{spinup}. On the other hand, the so called gravitational
radiation induced Rossby(r)-mode instabilities \cite{rmode} are
now believed to play a crucial role in setting an upper limit to
the spin  period \cite{Arras}. As a matter of fact, the
distribution of periods of all known pulsars drops off quite
sharply at about $2$~msec. All sub-millisecond pulsar search
campaigns made so far have been indeed  negative \cite{submill1},
\cite{submill2}. Under such circumstances the possible direct
observation of gravitational radiation from pulsars above $2$~KHz
appears to be extremely difficult, though it cannot be completely
ruled out.

\subsection{\label{sec:stoch}Stochastic Background}
The standard procedure to search for a
stochastic background of gravitational waves (unresolved
superpositions of gravitational wave signals of astrophysical or
cosmological origin) is to cross-correlate (and suitably filter)
the data of two (or more) detectors\footnote{
%--------------------------------
The power of a statistical hypothesis testing strategy based on a
single detector's output to decide whether any excess noise could
be interpreted as due to such a stochastic background would be
exceedingly poor \cite{Christ}.}
%--------------------------------
%--------------------------------
%\footnote{
%In this connection multi-mode acoustic detectors (e.g., spheres, \cite{LoboMon})
%and Michelson-Sagnac interferometers (e.g., LISA \cite{Cornish})
%might offer distinct advantages.}
%--------------------------------

Specific strategies for GW background detection using two bar
detectors \cite{twobars}, two spheres \cite{twoballs}, two
interferometers \cite{twoligos}, a bar and an interferometer
\cite{mixed} have been discussed by various Authors, and are in
current use to set upper limits on the gravitational wave
background power spectral density on the basis of available data
from operational gravitational wave antennas
\cite{Whelan}--\cite{Abbott}.

We plan to develop a similar (straightforward) analysis for the
special relevant cases of two MAGOs, one MAGO and one bar (or
sphere), and one MAGO and one interferometer.

In view of the easy tunability of MAGOs at any frequency in the
range between $10^3$~Hz and $10^4$~Hz and beyond, using MAGOs
might be interesting for probing the otherwise unaccessible high
frequency part of the sought stochastic gravitational wave
spectrum.

\section{Detector Layout}
\subsection{Electromagnetic design} In order to build an efficient
detector, a suitable cavity shape has to be chosen. According to
some general considerations, a detector based on two coupled
spherical cavites looks very promising (Fig. \ref{fig:paco2})
\cite{rfsc01}. The choice of the spherical geometry is based on
several factors. From the point of view of the electromagnetic
design the spherical cell has the highest geometric factor $G$,
thus it has the highest electromagnetic quality factor $\cal{Q}$,
for a given surface resistance $R_s$ (${\cal{Q}} = G/R_s$). For
the TE$_{011}$ mode of a sphere, the geometric factor has a value
$G \approx 850 \, \Omega$, while for standard elliptical
radio-frequency cavities used in particle accelerators, the
TM$_{010}$ mode has a value $G \approx 250 \, \Omega$. Looking at
the best reported values of surface resistance of superconducting
accelerating cavities, which typically are in the
$10^{-8}\,\Omega$ range, we can extrapolate that the
electromagnetic quality factor of the TE$_{011}$ mode of a
spherical superconducting cavity can be ${\cal{Q}} \approx
10^{10}-10^{11}$.

In the first generation of detectors, dedicated to the development
of the experimental technique, the internal radius of the
spherical cavity will be $r \approx 100$~mm, corresponding to a
frequency of the TE$_{011}$ mode $\omega \approx 2$~GHz. The
overall system mass and length will be $M \approx 5$~kg (with a
wall thickness $w\approx 2$~mm) and $L \approx 0.5$~m. The choice
of the wall thickness is made considering both practical and
design constraints. On one hand, the wall thickness should be kept
small enough to allow an easy fabrication while maintaining
sufficient stiffness to withstand the external pressure once the
cavity is evacuated. Furthermore, wall thickness was chosen to
optimize the cavity cooling process, and to guarantee optimum
stability against point--like thermal dissipation due to possible
defects present on the cavity inner surface. On the other hand,
wall thickness can be used to design a particular mechanical
resonant frequency and it is obviously related to the mass of the
detector, which plays an important role in the signal to noise
ratio.

Since this type of detector is ideally suited to explore the high
frequency region of the g.w. spectrum, we plan to build a tunable
cavity with $4$~kHz$\leq \oma - \oms \leq 10$~kHz, which is
outside the spectral region covered by the resonant and
interferometric detectors, both existing and planned, and is still
in a frequency region where interesting dynamical mechanisms
producing g.w. emission are predicted \cite{schutz2003,
grqc0109054, kokko2}.

\begin{figure}[hbt]
 \begin{center}
 \includegraphics[scale=0.4]{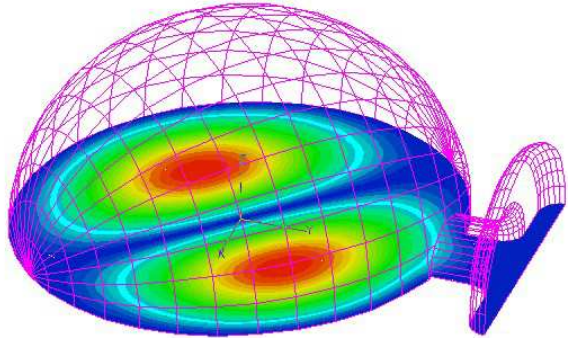}
 \caption{
 \label{fig:magexy}
 Electric field magnitude of the TE$_{011}$ mode. Note the alignment of the field axis}
 \end{center}
 \end{figure}

The interaction between the g.w. and the detector is characterized
by a transfer of energy and angular momentum. Since the helicity
of the g.w. (the angular momentum along the direction of
propagation) is $2$, the g.w. can induce a transition between the
two levels provided their angular momenta differ by $2$; this can
be achieved by putting the two cavities at right angle or by a
suitable polarization of the electromagnetic field inside the
resonator. The spherical cells can be easily deformed in order to
induce the field polarization suitable for g.w. detection. The
optimal field spatial distribution has the field axes in the two
cavities which are orthogonal to each other (Fig.
\ref{fig:magexy}). Different spatial distributions of the e.m.
field (e.g. with the field axes along the resonators' axes) have a
smaller effect or no effect at all.

\begin{figure}[hbt]
 \begin{center}
 \includegraphics[scale=0.7]{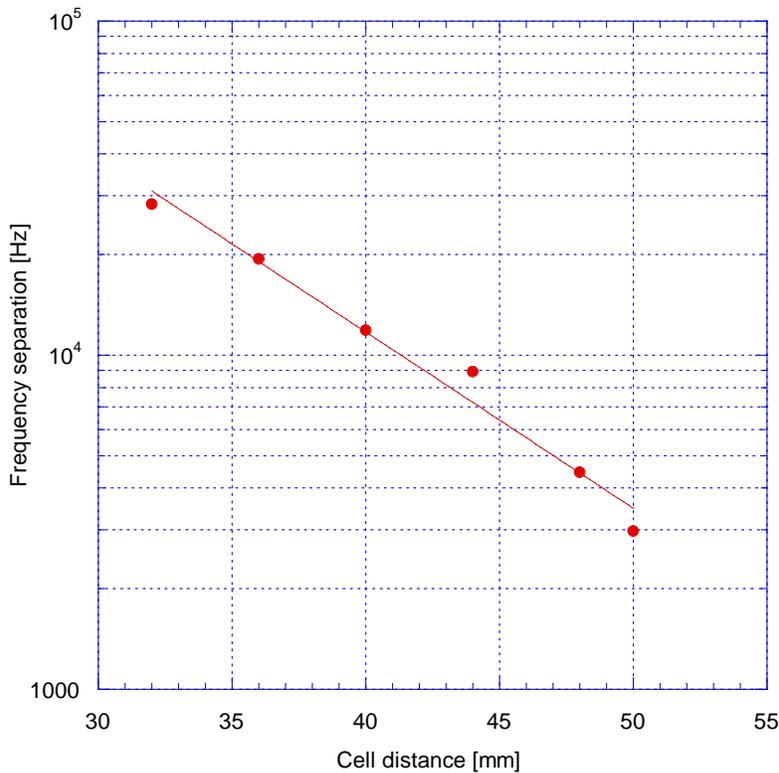}
 \caption{\label{fig:tuning}
 Detection frequency vs. coupled cells distance}
 \end{center}
 \end{figure}

\begin{figure}[hbt]
 \begin{center}
 \includegraphics[scale=0.7]{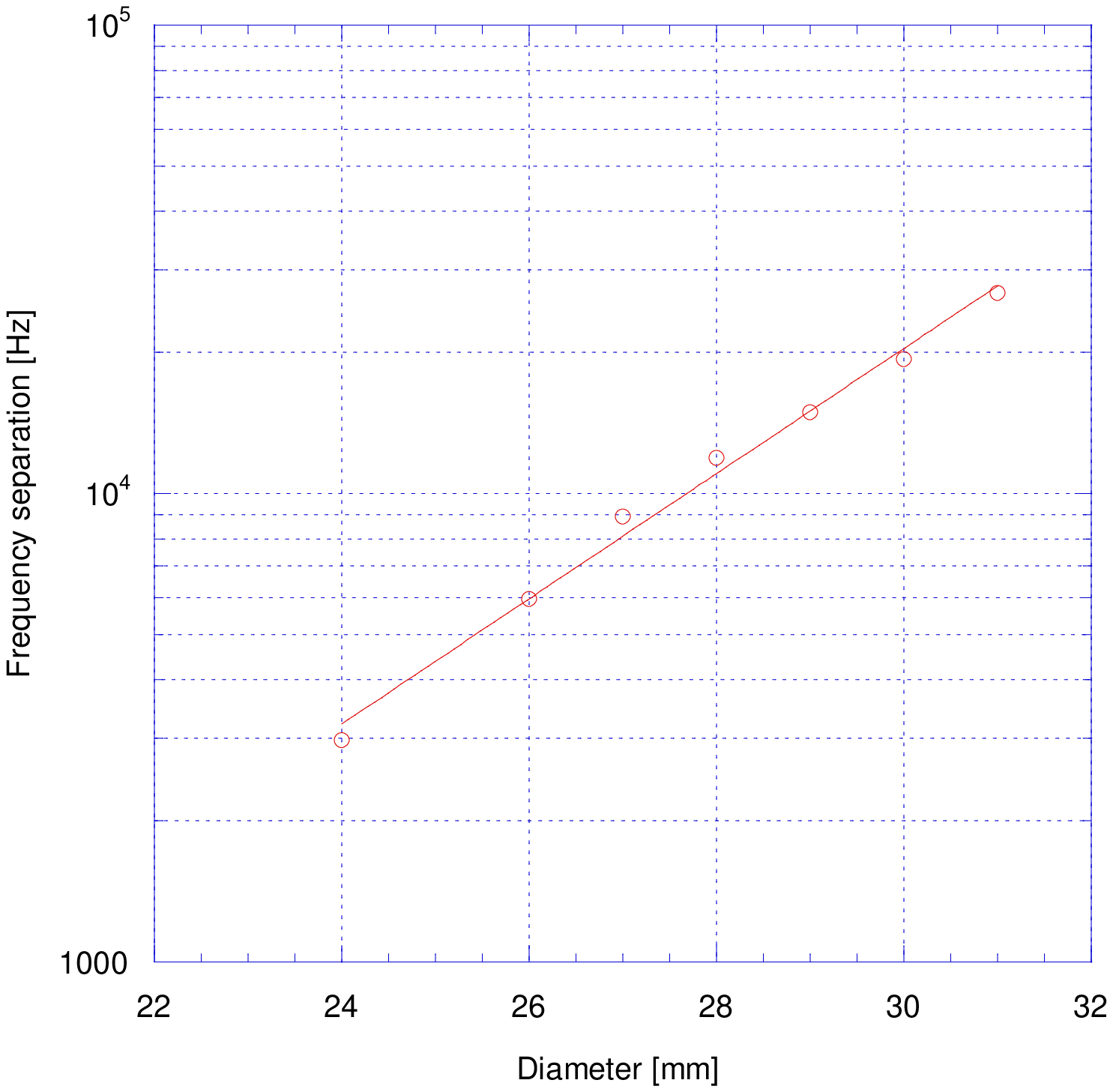}
 \caption{\label{fig:tuning2}
 Detection frequency vs. coupling tubes diameter}
 \end{center}
 \end{figure}

A tuning cell is inserted in the coupling tube between the two
cavities, allowing to tune the coupling strength (i.e. the
detection frequency) around the design value. The dependence of
the detection frequency on the distance between the two coupled
cells is shown in Fig. \ref{fig:tuning}, while its dependence on
the diameter of the tubes is shown in Fig. \ref{fig:tuning2}.

One detector based on two spherical niobium cavities (with fixed
coupling) has recently been built and tested at CERN. A second detector with variable coupling has
also been built and is now being tested (Fig. \ref{fig:tunable}).

\begin{figure}[hbt]
 \begin{center}
 \includegraphics[scale=0.7]{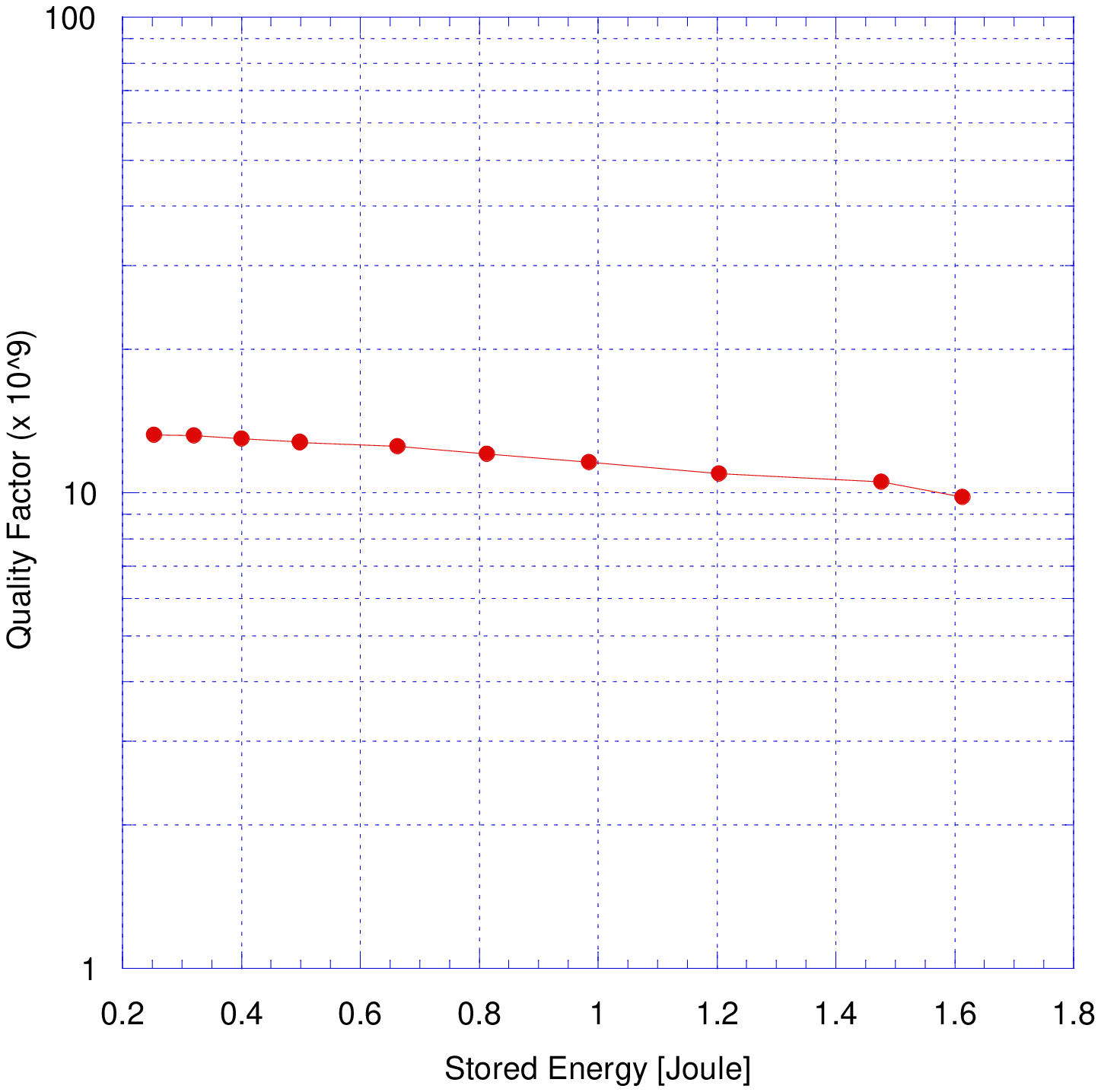}
 \caption{
 \label{fig:misuraq}
 Quality factor vs. stored energy for the fixed--coupling cavity
 }
 \end{center}
 \end{figure}

The first test on the cavity with fixed coupling, showed a
quality factor ${\cal{Q}} \geq 10^{10}$ (see Fig.
\ref{fig:misuraq}). This corresponds to a surface resistance
$R_s\approx 50$~n$\Omega$, a factor of ten higher than the best
values reported for superconducting accelerating cavities. The
obtained result is very satisfactory. In fact, the whole
fabrication procedure (including surface treatments) is optimized
for the elliptical cavity geometry used for high energy particle
acceleration. Some development is still needed to tailor the
technique to the spherical shape of our resonator and to obtain a
surface quality comparable to that routinely obtained on
elliptical cavities that would lead to a quality factor ${\cal{Q}}
\approx 10^{11}$.

\subsection{Mechanical design} From the mechanical point of view it
is well known that a spherical shell has the highest interaction
cross-section with a g.w. and that only the quadrupolar mechanical
modes of the sphere do interact with a gravitational wave
\cite{lobo}. The mechanical design is highly simplified if a
hollow spherical geometry is used. In this case the deformation of
the sphere is given by the superposition of just one or two normal
modes of vibration and thus can be easily modeled. In fact, the
proposed detector acts essentially as an electro--mechanical
transducer; the gravitational perturbation interacts with the
mechanical structure of the resonator, deforming it. The e.m.
field stored inside the resonator is affected by the time--varying
boundary conditions and a small quantity of energy is transferred
from an initially excited e.m. mode to the initially empty one. We
emphasize that our detector is sensitive to the polarization of
the incoming gravitational signal. Once the e.m. axis has been
chosen inside the resonator, a g.w with polarization axes along
the direction of the field, will drive the energy transfer between
the two modes of the cavity with maximum efficiency. With standard
choices for the axes and polar coordinates, the pattern function
of the detector is given by $F_\times = -\cos(\theta)\sin(2\phi)$,
and is equal to the pattern function of {\em one} mechanical mode
of a spherical resonator.

\subsection{Status and results on the simulation of the mechanical modes}
\label{sec:mechsim} We have run an ANSYS mechanical simulation of
our prototype over a wide frequency range ($0$--$7000$ Hz),
finding approximately 180 resonant modes and we have compared
these modes to the (analytical) modes of a spherical shell for
identification.

We foresee to suspend the detector by a region near its center of
mass, a region where the tuning system should also be implemented.
For this reason, the model used for the simulation consists of
half detector (single cell) fixed on the central pipe, near the
whole detector center of mass. Let us focus on two frequency
ranges: the modes below the quadrupole frequencies (ANSYS
calculated $\nu_{quad}=3646$~Hz, theoretical frequency is
$\nu_{quad}=3750$~Hz) and the modes above $\nu_{quad}$.

\begin{figure}[hbt]
 \begin{center}
\includegraphics[scale=0.6]{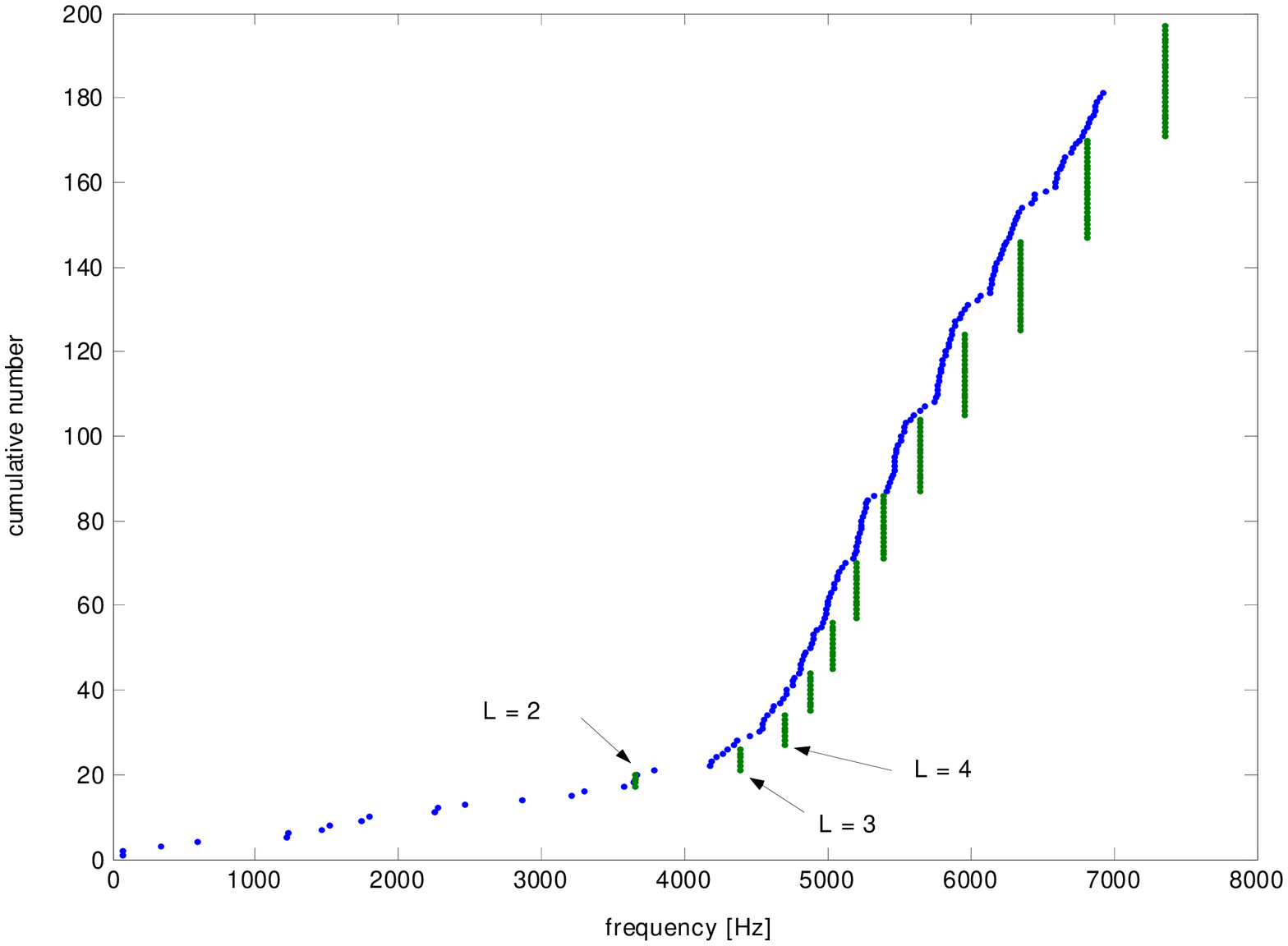}
 \caption{Spectrum of the mechanical normal modes of the \emph{small--scale}
 detector in the range $0 < f < 7000$~Hz.
 \label{fig:modiansys}
 }
 \end{center}
 \end{figure}

Below $\nu_{quad}$, we find a very limited number of modes ($\sim
16$), well separated in frequency. These modes are due to the
movement of the pipes with respect to the cavity (bending,
rotation and cantilever modes). Above $\nu_{quad}$ we find a great
number of modes, only related to the cell. Fig.
\ref{fig:modiansys} shows the cumulative number of modes versus
their frequency. The slope of the ANSYS calculation (blue dots) is
very close to the slope of the analytical spherical modes (green
cross), meaning that the number of the single cell modes found
above $\nu_{quad}$ corresponds to the ideal sphere, although
several mode shapes may be distorted due to the presence of the
pipes.

The frequency range around the quadrupolar mode of interest
($N$,$L$,$M$ = 1,2,2) gives us two main hints:
\begin{itemize}
    \item The rightfulness of the detection mode, which is minimally
    perturbed by the pipes attached to the cavity and oscillates
    approximately at the wanted frequency;
    \item The presence of only quadrupolar modes in the range of interest.
\end{itemize}

The first quadrupolar mode resembles one of the fundamental quadrupole mode of a
single, ideal sphere. The five quadrupolar modes are all near to
each other, although the cavity pipes interference gives a
discernible spread. A visual identification with the ideal
spherical modes can be done for the $L=2$, $L=3$ and $L=4$ mode
groups. Higher order modes, although somewhat similar to the ideal
case, are in general less easy to identify.

\subsubsection{Role of other mechanical mode scarce}
Consider now a detector tuned to an angular frequency $\Omega =
\oma-\oms$ and let us focus on OMR operation, so that $\Omega >
\omega_{mech}$. The signal, driven by the gravitational wave,
comes from the quadrupolar modes (in particular, for a well
oriented source, from the $N$,$L$,$M$ = 1,2,2 mode). That is, the
gravitational wave couples to those modes, driving them at angular
frequency $\Omega$. The thermal noise, which will surely come at
least from the same quadrupoles, may in principle come from other
mechanical modes. In particular, we seek deformations which
contribute to the noise but are not affected by the signal (g.w.).
For a mode to contribute to noise sources though, it must fulfill
two stringent requirements:
\begin{itemize}
\item[a)] The coupling coefficient
$C_m^{21}=\int_S \left(\vec H_2\cdot\vec H_1 - \vec E_2\cdot\vec
E_1\right)\vec\xi_m\cdot\,d\vec S$, between the electromagnetic
modes and the mechanical mode must not be null;
\item[b)] The mode frequency must be relatively near to the detection frequency.
\end{itemize}

The first requirement is the most important.

In the ideal, spherical case, the coupling coefficient $C_m^{21}$
is different from zero for the class of quadrupole modes alone.
Other angular numbers exhibits a strictly zero coefficient due to
symmetry mismatch between the electromagnetic field and the
mechanical mode.

A clue for this behavior comes from the chosen electromagnetic
mode angular dependence, which is $\propto \cos{(2 \theta)}$ (it
is not depending on $\phi$), while the radial component of the
displacement is proportional to $\cos{(L \theta)}$. This
calculation has been performed on an ideal, hollow sphere and
using analytical expressions for the electromagnetic fields. We
understand that these results must be checked against the more
realistic deformations delivered by ANSYS, and our future efforts
will surely tackle this problem. We are confident though that the
general behavior will be consistent, since the preliminary
analysis shows a considerable similarity between sphere modes and
the ANSYS model.

\section{Suspension System}
Like a resonant bar, the whole detector must be thoroughly
insulated from the environment and the mechanical design of the
insulation is therefore strongly bonded with the cryogenic design.
In particular, the insulation for the direct LHe supply to the
cavity (see section \ref{sec:cryo}) may prove to be a formidable
task. The mechanical insulation foreseen for this detector may be
inspired by the already existing resonant bars suspension systems,
although some major differences exist.

\begin{figure}[hbt]
 \begin{center}
 \includegraphics[scale=0.45]{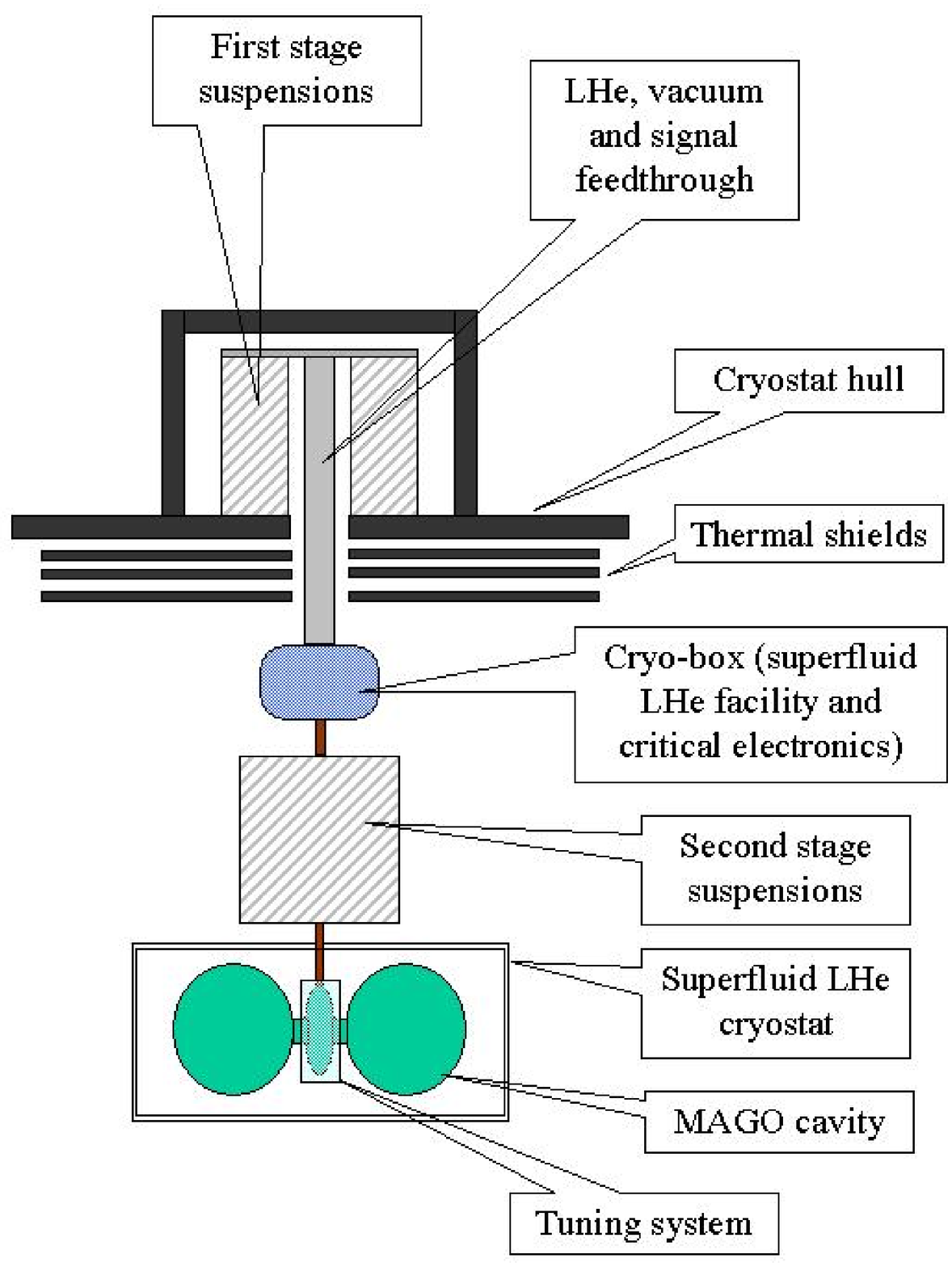}
 \caption{
 \label{fig:sosp}
 Sketch of the suspension system
 }
 \end{center}
 \end{figure}

Since one of the feature of this detector is the tunability, the
mechanical design should provide adequate insulation in the
frequency range 4--20 KHz and a very rough estimation suggests a
minimum value of $250$ dB of attenuation in the desired band (by
comparison, Al bars suspension are estimated to provide $\approx
300$ dB of insulation). A sketch of the suspension system is shown
in fig. \ref{fig:sosp}.

The room temperature filter bank is attached to the outer cryostat
hull and will provide a first decoupling to the environment. There
follows a cryogenic box, a facility which will host some critical
low-temperature electronic components and the source for the
superfluid LHe, and finally the cryogenic filter bank, which
should provide the required insulation. This second stage filter
is connected to the innermost cryostat, which is a superfluid LHe
container hosting the detector and its tuning system. The non
trivial task of filtering out environmental acoustical noise on a
wide frequency band can be more easily accomplished by optimizing
the two filter banks on separate frequency ranges, for example,
the first stage can be optimized for low frequencies. The
effectiveness of the cold filter (second stage) can be weakened by
the superfluid LHe feedthrough, which runs from the cryogenic box
to the MAGO cavity cryostat. Although the superfluid LHe can be
conveyed by means of a suitable capillary tube, acoustic
vibrations in LHe, triggered in cryo-box for instance, require a
deeper understanding and maybe a dedicated damping system. The
mechanical insulation design is made more complex by the MAGO
cavity plus the superfluid cryostat weight. Since their weight is
of the order of some tens of kilos, it is comparable to the
suspensions weight and therefore, the load on uppermost suspension
element can be considerably different from the lowermost one. All
these problems will be addressed in the following years, where a
comprehensive cryogenic and mechanical design will constitute the
main effort.

\subsection{Seismic Noise}
Preliminary seismic noise measures have been performed in our
laboratory during working hours (day time). The measures were
taken in a range from 10 to 9000 Hz, which should cover the
sensitive region for the MAGO detector. By comparing the data to
the expected level of the thermal noise we can estimate the
required attenuation of the external environment noise to be (at
least) of the order of 180--200 dB in the range 500 Hz--9 kHz, and
160 dB if we limit the frequency range to $f > 4$ kHz.

\subsubsection{Data taking}
Seismic noise was acquired with a PCB 393B12 accelerometer
connected to a Stanford research SR780 FFT analyzer. The
accelerometer has a nominal frequency range of 0.05 to 4000 Hz and
a sensitivity of 1019.4 mV/(m/s$^2$). Although that particular
accelerometer is not fit for measuring frequencies up to 9000 Hz,
we have nevertheless taken the frequency range 10--9000 Hz in
order to have a preliminary, order of magnitude measure of the
environmental disturbances in the range of interest. We are
justified in taking the high frequency measures because the
accelerometer response has a rising characteristic with increasing
frequency (ref. model 393B12 operating guide). It is thus expected
that the high frequency portion of the seismic noise power
spectral density might be overestimated. More accurate evaluation
of the seismic noise in the high frequency range ($f > 4000$ Hz)
will be done as soon as our group is provided with a suitable
accelerometer (for instance the PCB 307B).

The accelerometer was secured to a 25 kg lead brick, which allowed
us to easily sample the vibrations along the three orthogonal
directions in space, defining axis $z$ oriented perpendicular to
the floor, whereas $x$ and $y$ parallel to the room walls. For
each direction, a measure with the accelerometer switched off
(null measure) was performed in order to take into account the
SR780 FFT and cables intrinsic noise over the whole frequency
range.

The sub--cooled liquid helium environment, needed for MAGO
operation, is provided by means of a big mechanical pump, located
next to the laboratory. When switched on, this pump emits an
audible noise, and for this reason measures were taken with the
pump both switched on and off.

The Volts measured as the output of the accelerometers were
converted into the equivalent force acting on our detector. This
allows direct comparison with the Langevin force spectral density
and therefore yields immediately the required attenuation.
Sticking to a conservative approach, we have assumed that the
noises measured in the three directions equally couple to the
detector.

\subsubsection{Results}
\begin{figure}[hbt]
\begin{center}
\includegraphics[scale=0.75]{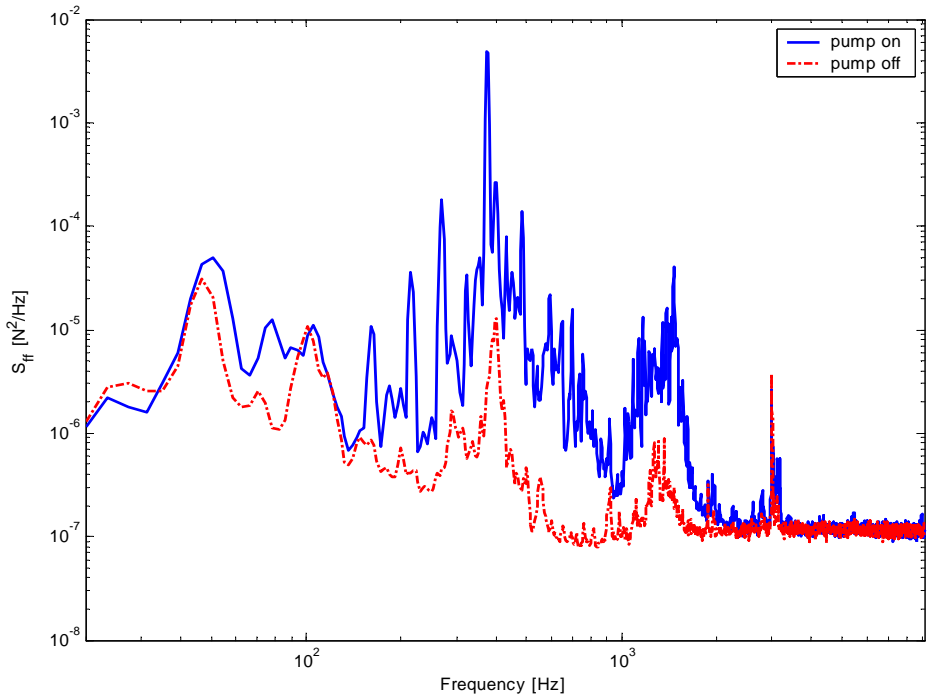}
\caption{Total force spectral density measured in the "quiet"
condition (helium pump switched off, dotted line) and "noisy"
condition (helium pump switched on, continuous line).
\label{fig:fig1a}}
\end{center}
\end{figure}

Figure \ref{fig:fig1a} shows the total force spectral density
(summed over $x$, $y$ and $z$) with the helium pump switched on
and off. The contribution of the pump is relevant up to a
frequency $f \sim 3$ kHz.

\begin{figure}[hbt]
\begin{center}
\includegraphics[scale=0.75]{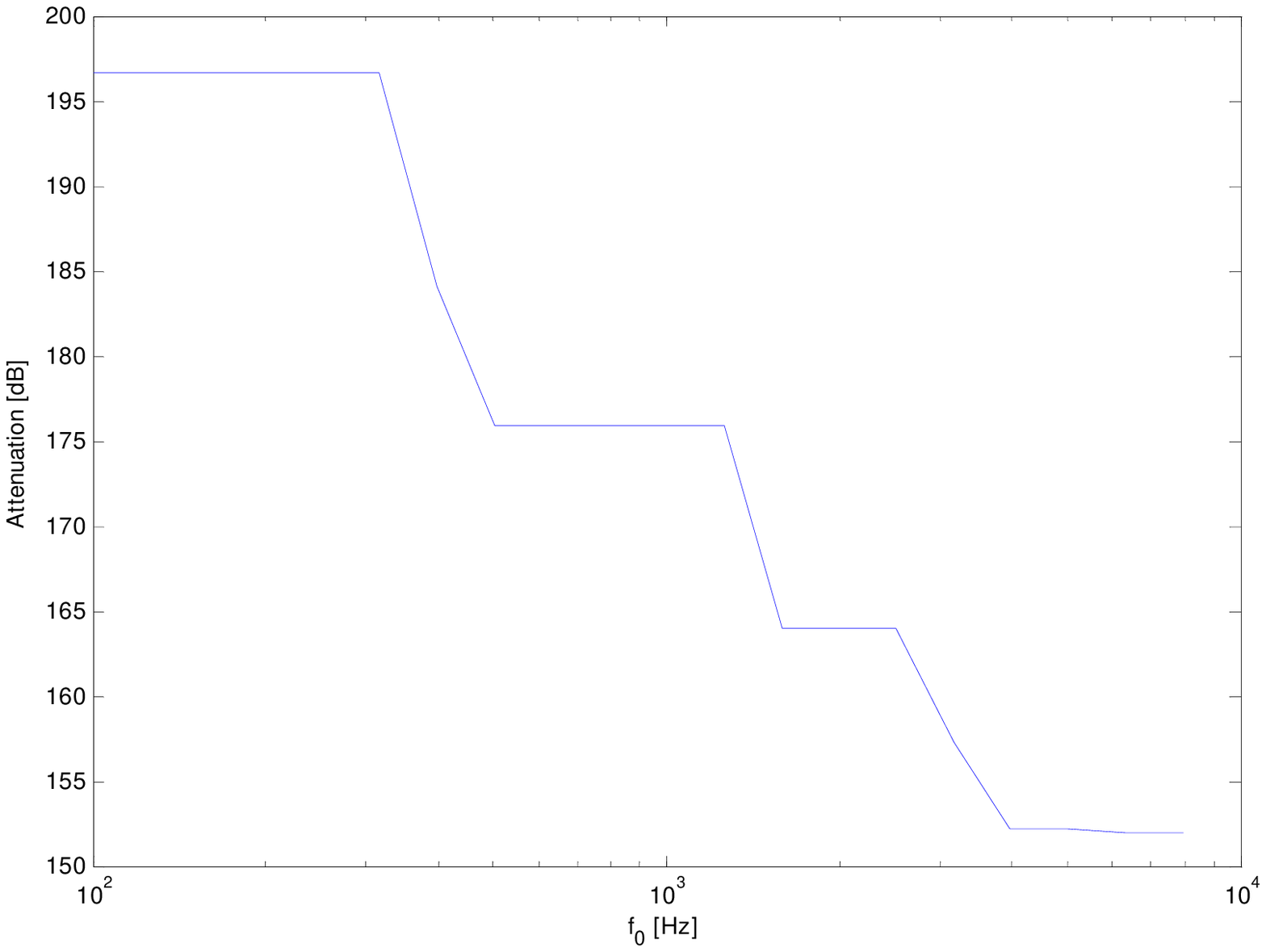}
\caption{Minimum attenuation as function of the cut--off frequency
$f_0$. \label{fig:fig2a}}
\end{center}
\end{figure}

If we require that the measured force spectral density be of the
same order of the Langevin force spectral density $S_{ff} = 4 K_b
T M \omega_m/Q_m$ , we immediately can calculate the approximate
attenuation for the range of frequencies greater than a chosen cut
off frequency $f > f_0$ (see fig. \ref{fig:fig2a}).

Although these measures are not conclusive, we expect the
attenuation values proposed in this document to be a fairly good
approximation. More accurate and reliable numbers could be
provided with suitable test equipments and a setup nearer to a
g.w. detector experiment. Note that the assumption of equal
coupling between the vibrations in the three directions and the
system may well be an overestimation of the real case. Further
studies could therefore relax the attenuation requirements.

\section{Cryogenics}
\label{sec:cryo} Assuming the conservative value of 1~W/m$^2$ for
the super--insulation radiation losses, the radiating heat load is
$P_{rad} \approx 4$~W. Adding to this figure a couple of watts to
account for the conduction losses in the cavity suspension
(tie--rods), we obtain the value of $P_{tot}\approx 6$~W as an
upper limit. These steady state losses have to be compared with
the RF cryogenic losses of our detector.

Assuming, in operation, a peak surface magnetic field $H_{surf} =
0.1$~T (half of the critical field of niobium) the stored energy
in the cavity will be $U \approx 30$~J. We remark that $H_{surf} =
0.1$~T corresponds (in an accelerating cavity) to 25 MV/m of
accelerating field, a value routinely achieved in the cavities
developed for the Tesla project. Given the geometric factor of our
cavity, $G =$ 835 $\Omega$, and assuming a residual surface
resistance of 5 n$\Omega$ (the surface resistance routinely
obtained in the LEP--II cavities at 1.8 K), we can foresee a
quality factor $\qst \approx 10^{11}$, corresponding to a
dissipation of 4 W in the helium bath (@ $H_{surf}=0.1$~T).

The planned operating temperature of our system is 1.8 K to fully
exploit the advantages of the RF superconductivity. At the chosen
frequency of 2 GHz the surface resistance at 1.8 K is well
saturated at the residual value, avoiding the change of surface
resistance produced by the heating of the surface. The use of
superfluid helium as refrigerant guarantees a very good thermal
dissipation for the cavity, with an even temperature distribution
along the cavity surface.

\begin{figure}[hbt]
 \begin{center}
 \includegraphics[scale=1]{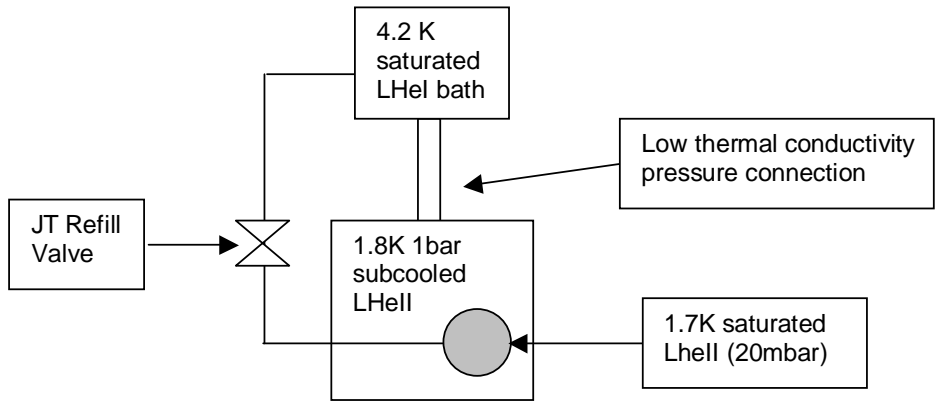}
 \caption{
 \label{fig:refrig}
 Flow chart of the subcooled superfluid helium refrigerator
 }
 \end{center}
 \end{figure}

To further improve the heat exchange and reduce the effect of the
helium boil--off we propose a scheme of refrigeration similar to
the one foreseen for the LHC magnets and already used since the
eighties for the refrigeration of high field magnets (e.g.
TORE--II supra). The flow chart of the refrigerator is shown in
figure \ref{fig:refrig}.

This refrigeration scheme, using subcooled superfluid helium, at
atmospheric pressure and 1.7--1.8 K, avoids all the problems
related to the bubbles in the refrigerating bath, improves the
Kapitza resistance at the helium--cavity wall interface and
increases by 30\% the peak transfer value to the helium bath. This
refrigeration scheme will be the best solution also from the point
of view of the mechanical noise induced on the detector by the LHe
bath. The subcooled operation at 1.8 K and atmospheric pressure
(compared with saturated helium at 1.8 K and 20 mbar), will
eliminate the possibility of the bubble gas production. This
effect of an abrupt reduction of the bath induced mechanical noise
below the helium $\lambda$--point was already experienced on Weber
type gravitational detectors \cite{fulvioricci}.

To form a bubble of a certain size in subcooled helium we need not
only to transfer to the helium bath the amount of heat
corresponding to the heat of vaporization, but also the enthalpy
needed to reach the normal boiling point of the gas at 4.2 K.
Furthermore the Claudet--type refrigerator reduces to a minimum
the refrigerator--bath interaction minimizing the mechanical noise
coming from the pumps and from the saturated HeI bath. The
sub--atmospheric HeII bath cools the main bath only by conduction
via the heat exchanger HT, the refill of this bath is done trough
the JT needle valve working on the liquid helium flow. The
hydraulic impedance of this needle is quite high and damps any
pressure fluctuation due to the pumping system. The 4.2 K bath is
in hydraulic contact with the 1.8 K sub--cooled bath via the high
impedance duct to reduce the heat input to the superfluid helium
and allows to reduce the bath temperature well below the
$\lambda$--point of the helium (2.19 K). To obtain this result we
need a quite long and narrow hydraulic channel with a very high
thermal impedance, but also the hydraulic impedance of the channel
is high, decoupling the superfluid bath from the saturated (4.2 K)
or nearly superfluid bath. In this way pressure fluctuations due
to the He bath refilling or to turbulences, produced by
thermo--acoustic oscillations of the saturated bath, will have a
negligible effect on the superfluid bath.

Coming to the conclusions the proposed refrigeration scheme will
greatly help in reducing the sensitivity of the proposed detector
to the acoustic noise produced by the fluctuation of the helium
bath wile giving us the more comfortable situation by the point of
view of the RF performances.

\subsection{Refrigeration scheme and Noise Issues}
%\subsection{General remarks}
The microwave power dissipation in the cavity walls is easily
found by the quality factor definition \beq Q=\frac{\omega U}{P}
\eeq with $\omega$ angular frequency of the radio--frequency (RF)
field, $U$ Electromagnetic energy stored in the resonator, and $P$
RF power lost in the cavity wall.

In our case (operating frequency $\omega\sim 2$~GHz) for a stored
energy $U\sim 18$~J, corresponding to cavity operation at peak
surface magnetic field  $B_{peak}\sim 100$~mT (already achieved on
the prototype cavity measurements), the power dissipation is
$P\sim 20$~W at the design electromagnetic quality factor
$Q=10^{10}$. The \emph{measured} quality factor for the prototype
MAGO cavity is $Q=5\times10^{10}$ at the design field, giving a
safety factor of \emph{five} over the power dissipation level
quoted in the proposal.

The RF power dissipation is fairly constant on more than 70\% of
the resonator surface; moreover the temperature dependence of the
RF surface resistance of the niobium is highly non linear
\cite{bardeen58}, the thermal conductivity of the superconductor
is very poor, and the heat capacitance low \cite[chapter
15]{zemansky}. For those reasons (for the quoted quality factor
and field) the resonator can only be cooled via an helium bath.

Localized thermal connections to an heat sink at low temperature,
as the one used for cooling bars, lead to thermal runaway of the
resonator over the critical temperature of the niobium, up to 20
to 30~K. Considerations on the thermal noise of the detector, and
the temperature dependence of the RF surface resistance of
superconductors suggest to operate the detector at the lowest
possible temperature. Practical considerations on readily
available refrigerators with cooling power exceeding 10 watt,
restrict our choice to a refrigerator scheme using superfluid
helium operating around $T=1.5$--$1.8$~K.

At the operating temperature of $T = 1.8$~K the largest practical
superfluid helium refrigerators, designed for the operation of
superconducting cavities, have a cooling power in the multi kW
range and superfluid helium refrigerators developed for
superconducting magnet operation have $\sim 10$~kW of
refrigerating power \cite{claudet88a}. The world largest dilution
refrigerator, the Compass refrigerator at CERN, has a maximum
cooling power of 1~W at 0.3~K and 10~mW at 80~mK \cite{berglund}.
Finally, considerations of sec. \ref{sec:numeric} on the impact of
the coolant fluid on the mechanical quality factor of the
resonator, strongly suggest to use as coolant superfluid helium at
$T\sim 1.6$--$1.5$~K.

\subsubsection{Refrigerator Scheme}
\begin{figure}[hbt]
 \begin{center}
\includegraphics[scale=0.5]{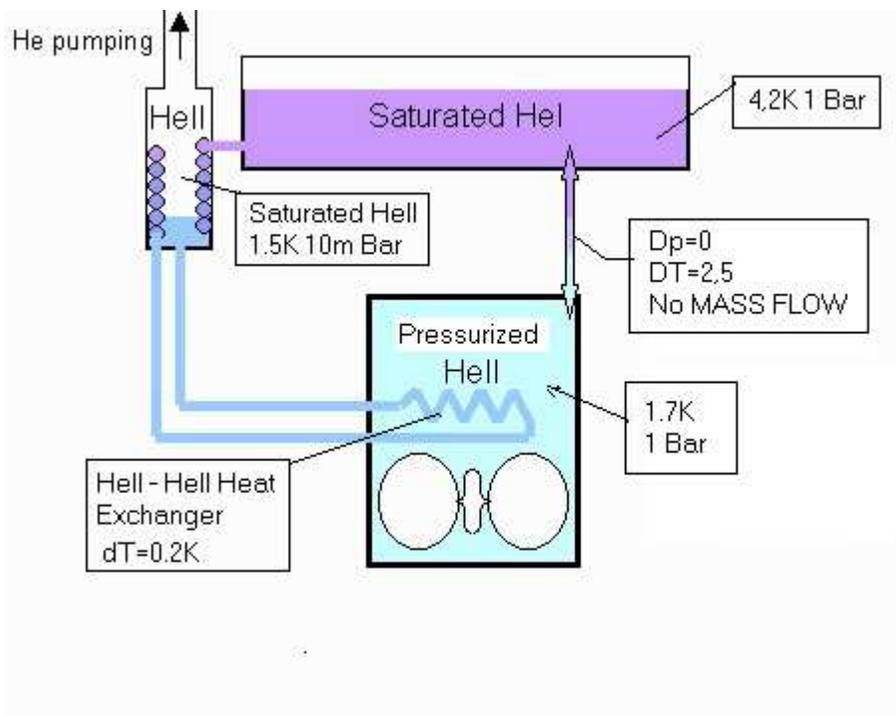}
 \caption{Pressurized Superfluid Helium (Claudet type) refrigerator.
 \label{fig:claudet}
 }
 \end{center}
 \end{figure}

The proposed refrigeration system is a pressurized superfluid
helium bath operating at $T=1.5$--$1.7$~K and $P=10^5$~Pa, based
on the original Claudet system developed at CEA for the \emph{TORE
II Supra} superconducting Tokamak \cite{claudet88b}. A conceptual
design of the refrigeration system is shown in fig.
\ref{fig:claudet}.

The cavity is immersed in a pool of stagnant superfluid helium and
is cooled by \emph{conduction} (not convection) by the surrounding
pressurized superfluid helium bath at $T=1.7$~K and $P=10^5$~Pa.

The operation of the cryostat is the following: the reservoir
(housing the cavity) is filled with liquid superfluid helium at
one bar. The superfluid helium in the reservoir is kept at
$T=1.7$~K by the HeII--HeII heat exchanger cooled at 1.5~K by the
saturated superfluid helium produced by pumping on the saturated
HeII pot. The pressure on the reservoir is kept at one bar by the
hydraulic connection (a narrow channel) with the saturated liquid
helium (HeI) at one bar and 4.2 K. Inside this narrow channel lies
the HeI--HeII interface with a temperature drop of 2.5~K, and no
pressure drop because no liquid helium flow takes place in the
channel.

The heat produced by the RF power dissipation on the cavity walls
is transferred to the HeII--HeII heat exchanger by the large
thermal conductivity of the superfluid helium. This large thermal
conductivity, (together with the peculiar properties of the
superfluid helium) keeps homogeneous the temperature of the
refrigerant in the reservoir, preventing any convection in the
liquid and the generation of any flow--induced noise in the
reservoir. Moreover at the rated dissipation of 20~W, the power
flow per unit area is roughly $10$~mW(cm)$^{-2}$, a factor of 100
lower than the value for the onset of bubble formation in the
helium bath at the cavity surface. Roughly speaking, to produce
bubbles in pressurized HeII one needs to transfer to the bath the
extra energy needed to heat the liquid from 1.7 K to 4.2 K.

Both effects, tightly connected to the use of pressurized
superfluid helium, will guarantee us against noise generation in
the process of cavity refrigeration.

\subsubsection{Induced noise in the cavity reservoir}
The proposed refrigeration scheme is quite safe in keeping to a
minimum the noise generated from the thermal dissipation in the
liquid helium bath. An additional source of acoustic noise is the
pressure fluctuation in the helium reservoir.

Two main sources inducing acoustic waves in the helium are the
pressure fluctuations on the saturated 4.2~K helium bath and
pressure fluctuations induced by the pumping system on the
saturated 1.5~K HeII pot feeding the HeII--HeII heat exchanger.

The pump--related pressure fluctuations on the HeII pot are
decoupled from the pressurized HeII reservoir by the metallic wall
of the heat exchanger. The two sections of the refrigeration
system are only in thermal contact, but mechanically separated.
Usually, the heat exchanger is a metallic box filled by the
superfluid helium produced in the pot. Moreover the thermodynamic
properties of the superfluid helium allow to relocate the
saturated HeII pot far away and to feed the heat exchanger by
mechanically decoupled lines (in the TORE Supra system the 1.5~K
pot is roughly 10 meters apart from the heat exchanger)
\cite{claudet88b}.

The pressure fluctuation on the 4.2~K bath are also rejected by
the system. The pressure connection between the pressurized
superfluid reservoir and the saturated helium bath need to be kept
as small as possible to improve the refrigerator efficiency.

We remind that no liquid helium flows trough the connections, and
that the steady state heat input at the 4.2~K HeI--1.7~K HeII
interface is produced by thermal conduction in the helium. This
heat flow is proportional to the cross section of the interface:
the smaller the channel, the better is the efficiency of the
refrigerator. Usually [5] the pressure connection is a tiny tube
of about 1~mm cross section or less. Pressure fluctuations in the
saturated 4.2~K HeI bath are greatly reduced, at the superfluid
helium side by the large hydraulic impedance of this
channel\footnote{Usually, (as an example in the refrigerators used
for magnet tests) the tiny tube is not even used, the pressure
connection being given by a non helium tight mechanical separation
between the two reservoirs.}.

\begin{figure}[hbt]
 \begin{center}
\includegraphics[scale=0.6]{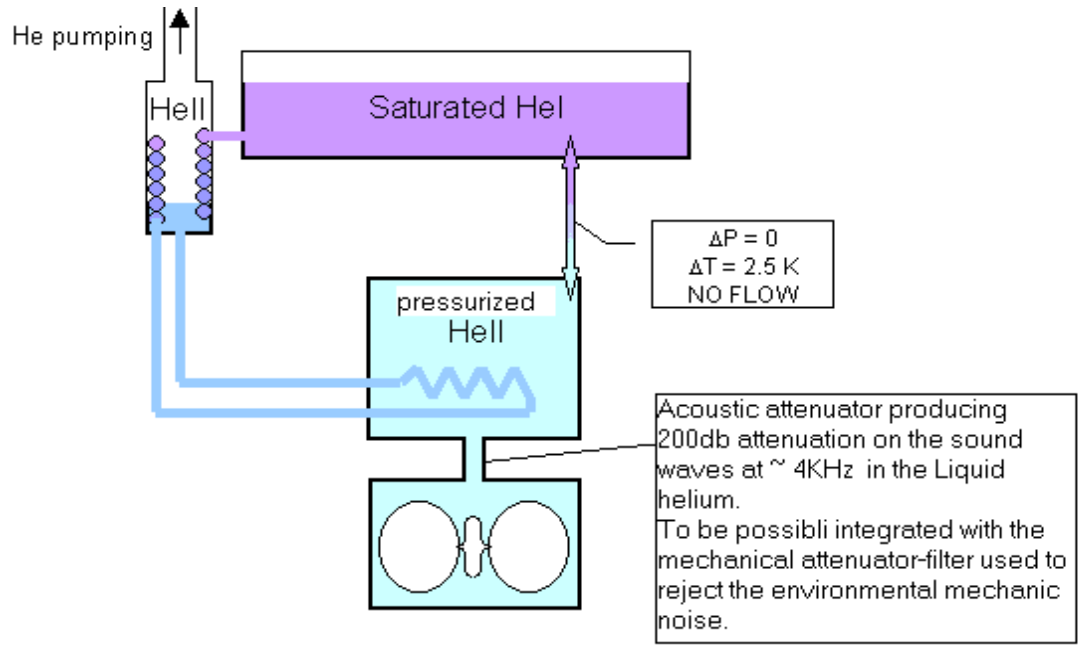}
 \caption{Modified refrigerator scheme including the hydraulic
 attenuator needed to decouple the detector from the mechanical
 noise converted to acoustic waves in the cooling bath.
 \label{fig:modif}
 }
 \end{center}
 \end{figure}

The last source of noise in the helium bath is the background
mechanical noise (seismic and man made) coupled to the liquid by
the motion of the walls of the refrigerator. By this way the
mechanical background noise can spoil the rejection of the
mechanical--suspension attenuator bypassing the filter via the
acoustic wave transmission of the liquid helium. To counter this
effect we foresee the modification of the refrigerator outlined in
fig. \ref{fig:modif}; the cavity helium II reservoir is
hydraulically decoupled from the upper part of the pressurized
HeII bath using an hydraulic low pass filter giving an attenuation
of 200~dB at 4~KHz. The needed attenuation can be obtained using a
4 cell low pass filter with a cut--off frequency of $\sim 4$~Hz,
producing an asymptotic attenuation of  $\sim 60$~dB per cell at
4~kHz.

In this way all the mechanical noise, converted to acoustic wave
in liquid helium, is confined in the upper reservoir. The upper
reservoir is also the starting point of the last stage of the
mechanical attenuator used to decouple the background mechanical
noise from the active part of the detector (the resonant RF
cavity) and can be (eventually) integrated with the mechanical
filter.

The heat transport in the superfluid helium being given by thermal
conduction, not by fluid convection, the only requirement on the
hydraulic filter is to guarantee a mean cross section of the
channel large enough to allow a temperature drop in the 0.1~K
range (worst figure).

\subsection{Mechanical dissipation in a liquid helium bath}
%\subsection{Absorption of sound}
\label{sec:absorption} The quality factor of a mechanical
resonator is given by the ratio of the elastic energy stored in
the mode of vibration and the power dissipated in one period: \beq
Q = \omega\frac{U}{P_{diss}}, \eeq where $\omega$ is the resonant
frequency. In vacuum, power can be dissipated only by the
intrinsic losses of the resonator ($P_{intr}$) and the resulting
\emph{intrinsic} quality factor is defined as $Q_0 = \omega
U/P_{intr}$. For niobium at low temperature and few kHz, $Q_0 \sim
10^{8}$.

If the resonator is immersed in a fluid, its oscillatory motion
causes a periodic compression and rarefaction of the fluid near
it, and thus produces sound waves. The energy carried away and
dissipated by these waves is supplied from the kinetic energy of
the body. Thus additional losses will be present and the quality
factor will be given by: \beq Q =
\omega\frac{U}{P_{intr}+P_{fluid}} \eeq

In a \emph{closed} vessel only \emph{stationary} sound waves can
exist. For such waves three main dissipation mechanisms can be
identified \cite{edmonds}:
\begin{enumerate}
    \item[1)] Power losses in the bulk of the fluid due to the
    classical processes of viscosity and thermal conduction:
    \beq
    \label{eq:pbulk}
    P_{bulk} = \frac{1}{2}\frac{\omega^2}{c^2}\left[\left(\frac{4}{3}\,\eta +
\zeta \right) + \kappa \left(\frac{1}{c_v} - \frac{1}{c_p} \right)
\right] \int_V u^2 \, dV
    \eeq
    \item[2)] Power losses in the viscous layer at the solid--fluid
    boundary, due to the tangential component of the fluid velocity at the boundary.
    The tangential fluid velocity should be zero at the wall, hence a tangential--velocity
    gradient must occur in the boundary layer of fluid, resulting in a viscous dissipation of energy:
    \beq
    \label{eq:pvisc}
    P_{visc} = \frac{1}{2}\left( \frac{\omega\rho_0\eta}{2}\right)^{1/2}\int_S (\vec u \cdot \vec n)^2 \, dS
    \eeq
    \item[3)] Power losses in the thermal layer at the solid--fluid boundary.
    In a sound wave, in fact, not only the density and the pressure, but also the temperature,
    undergo periodic oscillations about their mean values. Near a solid wall, therefore,
    there is a periodically fluctuating temperature difference between the fluid and the wall,
    even if the mean fluid temperature is equal to the wall temperature. At the wall itself,
    however, the temperature of the wall and the adjoining fluid must be the same.
    As a result a temperature gradient is formed in a thin boundary layer of fluid.
    The presence of temperature gradients, however, results in the dissipation of energy by thermal conduction:
\beq \label{eq:ptherm} P_{therm} = \frac{\rho_0^2c^2}{2p_0}\left(
\frac{\omega\kappa}{2\rho_0c_v}\right)^{1/2}\frac{\gamma-1}{\gamma}
\int_S u^2 \, dS \eeq
\end{enumerate}

\begin{table}
\begin{center}
\begin{tabular}{|cl|}
  \hline
  Symbol & Property name\\
  \hline
  $\rho_0$ & equilibrium density \\
  $p_0$ & equilibrium pressure \\
  $c$ & sound velocity \\
  $u$ & fluid velocity \\
  $\eta$ & shear viscosity \\
  $\zeta$ & kinematic viscosity\\
  $\kappa$ & thermal conductivity \\
  $c_v$ & specific heat at constant volume \\
  $c_p$ & specific heat at constant pressure \\
  $\gamma$ & specific heat ratio $c_p/c_v$ \\
  \hline
\end{tabular}
\caption{\label{tab:notation}Fluid properties and notation}
\end{center}
\end{table}

The calculation of the total energy dissipated into the fluid
requires the evaluation of the integrals over the fluid volume in
eq. \ref{eq:pbulk} and over the fluid--solid boundary in eq.
\ref{eq:pvisc} and eq. \ref{eq:ptherm}, which, in turn, depend on
the details of the fluid velocity pattern. In the following
section the calculation of the sound wave excited in a closed
spherical vessel by an oscillating spherical shell is done.

\subsubsection{Forced fluid motion in a spherical vessel}
\label{sec:model} 
\begin{figure}[hbt]
 \begin{center}
\includegraphics[scale=0.5]{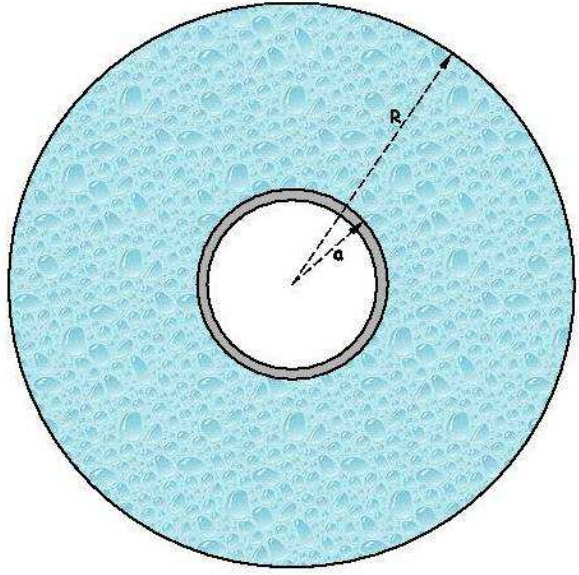}
 \caption{Sketch of the spherical shell (external radius $a$) enclosed
 in a concentric spherical vessel (radius $R$).
 \label{fig:bath}
 }
 \end{center}
 \end{figure}

The problem of the forced motion of a fluid in a
closed (rigid) vessel can be analytically solved only in simple
geometries. Here we shall focus on the problem of the calculation
of the sound wave excited by an oscillating spherical shell of
radius $a$ in a fluid which surrounds the shell and is enclosed in
a spherical vessel of radius $R$ (see. fig. \ref{fig:bath}).

A sound wave is completely described by a scalar function
$\phi(t,r,\theta,\phi)$ (the velocity potential) such that the
velocity field is given by $\vec u = \vec\nabla\phi$. The
mathematical problem can be casted in the following way: let us
find the scalar function $\phi(t,r,\theta,\phi)$ which satisfies
the wave equation (in spherical coordinates) \beq \label{eq:wave}
\nabla^2 \phi - \frac{1}{c^2}\frac{\partial^2\phi}{\partial t^2}=0
\eeq and the boundary conditions: \bear \label{eq:boundary}
(\vec\nabla\phi\cdot \vec n)|_a & = & u_0 f(\theta,\phi) \\
(\vec\nabla\phi\cdot \vec n)|_R & = & 0 \eear If we seek a
solution of the wave equation for the velocity potential which is
periodic in time, having the form: $\phi(t,r,\theta,\phi) =
\phi_0(r,\theta,\phi)\exp{(-i\omega t)}$, then we find for
$\phi_0$ the equation: \beq \label{eq:wave2} \nabla^2\phi_0 +
k^2\phi_0 = 0 \eeq where $k^2 = (\omega/c)^2$. The most general
solution of eq. \ref{eq:wave2} is given by the function \beq
\label{eq:gensol} \phi_0 = \sum_{l=0}^\infty\sum_{m=-l}^l R_l(r)
Y_l^m (\theta,\phi) \eeq where $Y_l^m (\theta,\phi)$ are the
spherical harmonics, and $R_l(r)$ satisfies the Bessel equation
\beq \label{eq:bessel} R_l'' + \frac{2}{r}R_l' + \left(k^2 -
\frac{n(n+1)}{r^2}\right)R_l = 0 \eeq which has solutions: \beq
\label{eq:bessol} R_l(r) = \frac{A_l J_1(l+\frac{1}{2},k r) + B_l
J_2(l+\frac{1}{2},k r)}{\sqrt{r}} \eeq where $J_1$ and $J_2$ are
the Bessel functions of the first and second kind.

If we suppose that the vibrating spherical shell oscillates with
quadrupolar pattern ($l=2$, $m=2$), so that the boundary condition
at $r=a$ can be put in the form $(\vec\nabla\phi\cdot \vec n)|_a =
u_0 Y_2^2(\theta,\phi)$ we get the following solution for the
sound wave induced in the fluid: \beq \label{eq:solution}
\phi_0(r,\theta,\phi) = \frac{A_2 J_1(\frac{5}{2},k r) + B_2
J_2(\frac{5}{2},k r)}{\sqrt{r}} Y_2^2 (\theta,\phi) \eeq with the
two constants $A_2$ and $B_2$ determined by the boundary
conditions.

\subsubsection{Numerical estimates}
\label{sec:numeric}
\begin{table}
\begin{center}
\begin{minipage}{100mm}
\begin{tabular}{|lll|}
  \hline
  Temperature & $T$ & $1.5$~K \\
  Pressure & $p_0$ & $10^{5}$~Pa \\
  Density & $\rho_0$ & $1.47\times10^2$~kg/(m$^3$) \\
  Shear viscosity & $\eta$ & $1.5\times10^{-6}$~kg/(m-sec) \\
  Kinematic viscosity\footnote{\cite[sec. 25]{putterman}} & $\zeta$ & $1.5\times10^{-5}$~kg/(m-sec) \\
  Thermal conductivity$^a$ & $\kappa$ & $0.1$~W/(m-K)\\
  Specific heat (const. vol.) & $c_v$ & $1150.3$~J/kg \\
  Specific heat (const. press.) & $c_p$ & $1151.1$~J/kg \\
  Specific heat ratio ($c_p/c_v$) & $\gamma$ & 1.000695 \\
  First sound velocity & $c$ & $242$~m/sec\\
  \hline
\end{tabular}
\end{minipage}
\caption{\label{tab:heprop}Physical properties of the liquid
helium bath}
\end{center}
\end{table}

Let us focus on a specific example. If we take the radius of the
vibrating spherical shell $a=0.1$ m and the radius of the external
vessel $R=0.3$ m, we get $Q\sim 10^6$, for a niobium sphere,
oscillating at $f = 4\times 10^3$ Hz (the frequency of the
fundamental quadrupolar mode), in a bath of subcooled, superfluid
helium at $T=1.5$ K and $P=10^5$ Pa. The physical properties used
in the calculation are summarized in table \ref{tab:propsmall}.
\begin{table}

\begin{center}
\begin{tabular}{|lll|}
  \hline
  External radius & $a$ & $0.102$~m \\
  Thickness & $w$ & $2\times10^{-3}$~m \\
  Density & $\rho_{Nb}$ & $8.6\times10^3$~kg/(m$^3$) \\
  Mass & $M$ & $2.2$~kg \\
  Angular frequency & $\omega$ & $2.5\times10^4$~rad/sec \\
  Intrinsic quality factor & $10^8$ & \\
  Vessel radius & $R$ & $0.3$~m \\
  Helium volume & $V$ & $0.1$~m$^3$ \\
  \hline
\end{tabular}
\caption{\label{tab:propsmall}Characteristics of the small--scale
system}
\end{center}
\end{table}

For a large system ($a=0.6$ m, $R=1$ m), oscillating in the
fundamental quadrupolar mode at $f=800$ Hz in a liquid helium bath
with the same characteristics, we obtain $Q\sim 10^7$. The
physical properties used in the calculation are summarized in
table \ref{tab:proplarge}.
\begin{table}
\begin{center}
\begin{tabular}{|lll|}
  \hline
  External radius & $a$ & $0.6$~m \\
  Thickness & $w$ & $0.2$~m \\
  Density & $\rho_{Nb}$ & $8.6\times10^3$~kg/(m$^3$) \\
  Mass & $M$ & $5.5\times10^3$~kg \\
  Angular frequency & $\omega$ & $5\times10^3$~rad/sec \\
  Intrinsic quality factor & $10^8$ & \\
  Vessel radius & $R$ & $1$~m \\
  Helium volume & $V$ & $3.3$~m$^3$ \\
  \hline
\end{tabular}
\caption{\label{tab:proplarge}Characteristics of the large--scale
system}
\end{center}
\end{table}

Obviously, in addition to the mechanism of absorption which have
already been considered, energy may also be lost from the sound
field in the following ways:
\begin{enumerate}
    \item by transmission of mechanical energy to the material of the vessel wall;
    \item by subsequent radiation of energy from the outer surface of the vessel
    to the surrounding medium; This loss might be expected to acquire significance
    at resonant frequencies of the vessel;
    \item by propagation of sound through the liquid helium tubing system.
\end{enumerate}
These items need to be studied in detail on a realistic model of
the whole cryogenic system and, at the end, will be the subject of
experimental check.

\section{Detection Electronics}
The three main functions of rf control and measurement system of
the experiment are:
\begin{enumerate}
\item{The first task of the system is to lock the rf frequency of the master
oscillator to the resonant frequency of the symmetric mode of the
cavity and to keep constant the energy stored in the mode. The
frequency lock of the master oscillator to the cavity mode is
necessary to fill in energy in the fundamental mode of the cavity.
The reduction of fluctuations of the stored energy to less than
100 ppm greatly reduces the possibility of ponderomotive effects
due to the radiation pressure of the electromagnetic field on the
cavity walls and helps to minimize the contribution of the
mechanical perturbations of the resonator to the output noise. The
frequency lock allows also to design a detection scheme
insensitive to fluctuations of the resonant frequency of the two
cavities forming our detector.}
\item{The second task is to increase the detector's sensitivity by driving the
coupled resonators purely in the symmetric mode and receiving only
the rf power up--converted to the antisymmetric mode by the
perturbation of the cavity walls. This goal can be obtained by
rejecting the signal at the symmetric mode frequency taking
advantage of the symmetries in the field distribution of the two
modes. Our system, despite of some additional complexity,
guarantees the following improvements over the one used in
previous experiments:
\begin{itemize}
\item{a better rejection of the phase noise of the master
oscillator obtained using the sharp resonance ($Q=10^{10}$) of the
resonator as a filter;}
\item{a better insulation of the drive and detector ports obtained by
using separate drive and detection arms of the rf system;}
\item{the possibility of an independent adjustment of the phase lag in
the two arms giving a better magic--tee insulation at the
operating frequencies;}
\item{a greater reliability for the frequency amplitude loop using the
transmitted power, instead of the reflected, coming from the
cavity.}
\end{itemize}
}
\item{The third task is the detection of the up--converted signal achieving the
detector sensitivity limit set by the contribution of the noise
sources \cite{bgpp2}. Slow pressure fluctuations on the cooling
bath, hydrostatic pressure variations due to the changes in the
helium level, pressure radiation, and so on, will change in the
same way the resonant frequency of both modes. Using a fraction of
the main oscillator power as local oscillator for the detection
mixer, the detection system becomes insensitive to frequency
drifts of both modes, allowing for a narrow band detection of the
up--converted signal produced by the cavity wall modulation.}
\end{enumerate}

\subsection{The rf control loop} \label{rf} The RF signal generated
by the master oscillator (HP4422B) is fed into the cavity through
a TWT amplifier giving a saturated output of 20 Watt in the
frequency range 2--4 GHz. The energy stored in the cavity is
adjusted at the operating level by controlling the output of the
master oscillator via the built--in variable attenuator.

The output signal is divided by a 3 dB power splitter. The $A$
output of the splitter is sent to the TWT amplifier, the $B$
output is sent, through the phase shifter (PS), to the local
oscillator (LO) input of a rf mixer acting as a phase detector
(PD). The output of the rf power amplifier is fed to the resonant
cavity through a

double directional coupler, and a $180^{\rm o}$ hybrid ring acting
as a magic--tee. The rf power enters the magic--tee via the sum
arm, $\Sigma$, and is split in two signals of same amplitude and
zero relative phase, coming out the tee co--linear arms 1 and 2.

The rf signal, reflected by the input ports of the cavity, enters
the magic--tee through the co--linear arms. The two signals are
added at the $\Sigma$ arm and sampled by the directional coupler
to give information about the energy stored in cavity allowing for
the measurement of the coupling factor, quality factor, stored
energy. While driving the cavity on the symmetric mode no
reflected signal is shown at the $\Delta$ port of the magic--tee
where the signals coming from the co--linear arms are
algebraically added to zero due to the $180^{\rm o}$ phase shift.

To get the maximum of the performances of the magic--tee we need
to have equal reflected signals (phase and amplitude) at the
cavity input ports. To preserve the signal integrity we use
matched lines (in phase and amplitude) inside the cryostat.
Because the phase shift is very sensitive to temperature
inhomogeneities between the two cables and the phase difference
between the two co--linear arms of the magic--tee gives a quite
strong signal at the $\Delta$ port, we need to compensate for
differential thermal contractions of the cables inside the
cryostat, leading to phase unbalance in the feed lines. To do that
we insert a phase shifter in one of the lines to compensate for
differences and to reduce to a minimum the leakage of the unwanted
modes on the two ports. As we will show in section
\ref{sec:sensit}, mode leakage of the symmetric mode to the
$\Delta$ port sets a limit to the system sensitivity increasing
the overall noise level of the detector.

Mode leakage of the antisymmetric mode to the $\Sigma$ port
reduces the system sensitivity by reducing the signal level
available for detection. The output ports of the cavity are
coupled for a maximum output signal on the antisymmetric mode
(detection mode) and the magic--tee is used to reject the rf power
at the frequency of the symmetric mode. A fraction of the signal
at the $\Sigma$ port is fed to the rf input of the phase detector
PD via a low noise rf amplifier. The intermediate frequency  (IF)
output of the phase detector PD is fed back to the rf master
oscillator to lock the output signal to the resonant frequency of
the resonator. The total phase shift around the loop is set
through the phase shifter PS, to have the maximum energy stored in
the detector. A careful design of the servo loop amplifier (SLA)
guarantees the stability of the system and the rejection of the
residual noise of the master oscillator up to one MHz. The same
fraction of the $\Sigma$ output of the output magic--tee is used
to keep constant, to 100 ppm, the energy stored in the cavity
feeding back an error signal to drive the electronically
controlled output attenuator of the master oscillator.

Great deal of care is needed in tailoring the frequency response
of both controls because the two loops can interact producing
phase--amplitude oscillations in the rf fields stored in the
cavities.

\subsection{Sensitivity enhancement using the mode symmetry}
\label{sec:sensit} The two modes of the detector cavity have (as
in the case of two coupled pendula) opposite symmetries of the
fields.

Using two separate sets of ports to drive the cavity and to
receive the up--converted signal at the frequency of the
antisymmetric mode, the cavity acts as a very sharp filter (due to
the high $Q$), with an high rejection of the noise coming from the
master oscillator at the frequency of the up--converted signal.
This already low residual noise, can be even more reduced in our
scheme using two magic--tees to drive the cavity purely in the
symmetric mode and to detect only the up--converted energy,
rejecting the unwanted field components by an amount given  by the
magic--tee insulation.

In the case of an ideal magic--tee the mode rejection is infinite.
If the cavity is driven purely in the symmetric mode no symmetric
mode component is transmitted through the system and there will be
no signal at the output port. In the ideal case this result is
obtained also in the more simple scheme used by Melissinos and
Reece \cite{rrm1,rrm2}, measuring the up--converted power coming
out of the detector along the input lines.

Our scheme gives better sensitivity and performances in the real
case. The first obvious gain is the sum of the $\Delta$ and
$\Sigma$ port insulation of the two tees, plus the possibility of
adjusting separately the input and output lines to get better mode
rejection. In a commercial magic--tee the insulation is specified
to be $\approx 25$ -- $35$ dB over its own bandwidth. The reason
for this quite low insulation is mainly due to the difficulty of
balancing on a large range of frequency the phases of the signals
coming from the two co--linear arms of the tee.

A phase unbalance as small as five degrees reduces the insulation
from $\Delta$ to $\Sigma$ port to only 25 dB.

Our electronic scheme allows for an independent compensation of
the magic--tee phase mismatch both at the feed frequency and at
the detection frequency in a flexible way: the phase mismatch is
compensated using a variable phase shifter at the input of one of
the co--linear arms. The optimum phase at the input side results
in a pure excitation of the symmetric (drive) mode, keeping the
power at the frequency of the antisymmetric (detection) mode 70 dB
below the level of the drive mode. Adjusting the phase at the
output will couple only the antisymmetric mode component,
rejecting the symmetric mode component by 70 dB. The total
symmetric--antisymmetric mode rejection of the system is the sum
of the attenuation we can obtain from the two $180^{\rm o}$
hybrids.

The input and output ports of the two cell cavity need to be
critically coupled ($\beta$ = 1) to the rf source and to the rf
detection system. In this way we have the optimum transfer of
power to the symmetric mode (a maximum of stored energy) and to
the antisymmetric mode (a maximum in the detector output). Because
the frequency and field distribution of the two modes are quite
close, the input and output ports are critically coupled to both
modes. For that reason 50$\%$ of the symmetric mode signal is
coupled to the idle $\Sigma$ port at the output magic--tee,

and symmetrically 50$\%$ of the antisymmetric mode signal is
coupled to the idle $\Delta$ port at the input magic--tee. {  We
remark that, since those two ports are not used in our detection
scheme, it is necessary to reflect back into the cavity the energy
flowing out of them. This task is achieved by a proper termination
of the ports; a careful analysis showed that closing the ports
with an open circuit completely decouples the input and output
arms (optimum rejection of the symmetric mode) and maximizes the
stored energy and the detector sensitivity.}

At the $\Sigma$ port of the detection arm we insert a directional
coupler to sample a tiny amount of the symmetric mode power coming
from the cavity. This signal is fed into the frequency--amplitude
servo loop used to lock the master oscillator to the cavity
frequency and to keep constant the energy stored in the cavity.

\subsection{Detection of the converted signal} Having devised a
means of suppressing the high level pump frequency signal, the
remaining problem is to amplify and detect the low level
parametric mode signal. The detection limit, hence overall
sensitivity of the machine is now determined by the RF thermal
noise level. In order to detect the low level signal it must be
amplified and to do this without degrading the signal to noise
ratio, a low noise amplifier is required. The lower the effective
noise figure of the RF amplifier, the higher will be the
sensitivity. In this case we can make use of the fact that the
resonant cavities are operated at cryogenic temperatures and use a
cryogenic low noise amplifier with lowest possible noise figure.

Since the late 1970s research has been conducted into the
performance of GaAs FET and later GaAs HEMT RF amplifiers at
cryogenic temperatures. The principle applications for this
technology have been for radio astronomy receivers and for
satellite ground station receivers. A typical receiver uses a 2
stage closed cycle Helium refrigerator reach a base temperature of
around 15 K. HEMT amplifiers work well at this temperature if
carefully design, but can also work down to liquid Helium
temperatures. However, the additional benefit in noise temperature
performance of operating at LHe compared to 15 K is not great and
is mostly not justified when traded-off against of system
complexity, when the equipment must operated in a radio astronomy
telescope or a satellite ground station antenna. However, in this
application, cooling to 4.5 K or 1.5K in liquid Helium is
available in the same dewar as the superconducting coupled
cavities.

\begin{figure}[hbt]
 \begin{center}
 \includegraphics[scale=0.7]{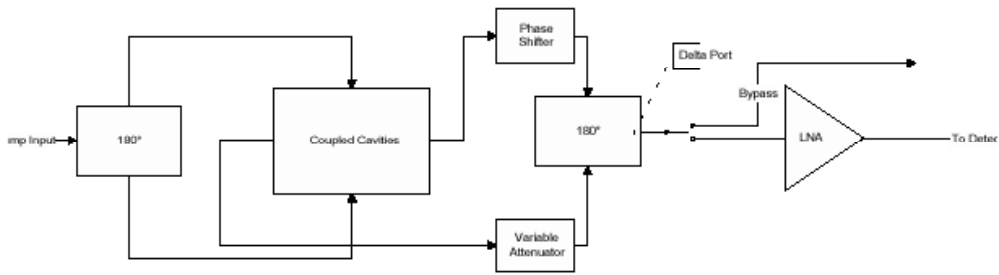}
 \caption{
 \label{fig:layout}
 Electric circuit of the detector system
 }
 \end{center}
 \end{figure}

The layout of the detection electronics is shown in fig.
\ref{fig:layout}.

The signal converted to the antisymmetric mode by the interaction
between the mechanical perturbation and the rf fields is coupled
to the $\Delta$ port of the detection arm of the rf system and
amplified by the low noise rf amplifier LNA.

The converted rf signal amplified by the LNA is fed to the rf
input port (RF) of a low noise double balanced mixer M1; the local
oscillator port (LO) of M1 is driven by the symmetric mode rf
power (at angular frequency $\omega_s$). The LO input level is
adjusted to minimize the noise contribution from the mixer. As
shown in the previous section the input spectrum to the RF port of
the mixer consists of two signals: the first at angular frequency
$\omega_s$ coming from the rf leakage of the symmetric mode
through the detection system; the second is the converted energy
at angular frequency $\omega_a$. Both signals are down--converted
by the M1 mixer giving to the IF port a DC signal proportional to
the symmetric mode leakage and a signal at angular frequency
$\Omega$ proportional to the antisymmetric mode excitation. The
down--converted IF output is further amplified using a low noise
audio preamplifier.

For the detection electronics of mechanically coupled
interactions, at angular frequency $\Omega$, as gravitational
waves, since the exact frequency and phase of the driving source
is not known, we can't perform a synchronous detection. We need to
perform an auto--correlation of the detector output,

or to cross--correlate the outputs of two different detectors.

The outlined detection scheme gives the benefit of being
insensitive to perturbations affecting in the same way the
frequency of the two modes.

The first point to note in terms of the general architecture of
the solution is that if a cryogenic amplifier is to be used within
the same cryogenic dewar as the coupled cavities, then the phase
shifter, variable attenuator and 180$^{\rm o}$ hybrid also need to
be inside the dewar. If this is not the case then the noise
temperature contribution of these components will dominate and
there will be little, if any, benefit in having a cooled
amplifier. As it is known that HEMT amplifiers can be produced for
cryogenic operation, we shall first focus on the passive
components. A consequence of placing the phase shifter and
variable attenuator inside the cryogenic dewar is that they will
have to be controllable by an electrical signal, rather than
manually. For this reason the description below considers voltage
controlled devices. In the interest of implementing a solution
which is reasonably economic, the preferred approach will be to
use commercial devices for the passive components, rather than
developing special--to--type components. Many electronic
components, particularly passive ones, will operate at cryogenic
temperatures, even if not specified to do so by their
manufacturers. Hence, the following focuses on the potential
problems which will need to be investigated in order to qualify
commercial components.

The 180$^{\rm o}$ hybrid is essentially a just a strip line
device. The only concern about operation in cryogenic temperatures
is due to differential thermal contraction of the materials of the
component, particularly in the area of the connections between the
strip line and coaxial I/O connectors. There may also be changes
in dielectric properties of the substrate, which may affect the
impedance of the I/O ports. Nevertheless strip line couplers have
been used in many cryogenic applications without any problem, so
we shall only validate this component with a simple test.

Voltage controlled phase shifter. The key element of the shift
shifter is a varactor diode. Silicon and GaAs diodes are know to
be capable of operation at cryogenic temperatures. However,
particularly in the case of silicon, the electrical
characteristics change significantly at cryogenic temperatures.
Therefore, it is predicted that a phase shifter will still operate
a cryogenic temperatures but its transfer characteristics will
change significantly from the ambient temperature specifications.
For this reason a particular aim will be to calibrate the device
when operating at cryogenic temperatures. Another factor to note
is that varactor diode phase shifters have limitations in
operating power. Between 0 dBm and + 5dBm input power the varactor
diodes begin to work as rectifiers. In any case, the pump energy
in the cavity output ports is typically below 0dBm, but this
should be confirmed.

Voltage controlled variable attenuator. The key element of a
voltage variable attenuator is a diode and so the analysis given
above for the phase shifter is also relevant for the variable
attenuator.

Bypass Switch. The function of the bypass switch is to allow
coarse tuning of the circuit for pump frequency suppression.
Before the phasing is tuned the pump frequency signal may be large
and could saturate the low noise amplifier. The bypass switch will
allow for direct connection of a spectrum analyser outside the
dewar for initial tuning. The bypass switch will be coaxial relay
type.

\section{Response of the Detector}
\subsection{Signal} The interaction between the stored e.m. field
and the time--varying boundary conditions depends on both how the
boundary is deformed by the external perturbation and on the
spatial distribution of the fields inside the resonator. Any field
configuration inside the resonator can be expressed as the
superposition of the electromagnetic normal modes \cite{slater}:
$\mathbf E(\mathbf r,t) = \sum \est_n(t) \, \mathbf E_n(\mathbf
r)$; $\mathbf H(\mathbf r,t) = \sum \hst_n(t) \, \mathbf
H_n(\mathbf r)$, with $\est_n(t) \equiv \sqrt{\epsilon_0} \int
\mathbf E \cdot \mathbf E_n \, dV$; $\hst_n(t) \equiv \sqrt{\mu_0}
\int \mathbf H \cdot \mathbf H_n \, dV$ and $\int \mathbf H_n
\cdot \mathbf H_m \, dV = \int \mathbf E_n \cdot \mathbf E_m \,
dV= \delta_{nm}$.

Similarly if $\mathbf u(\mathbf r,t)$ denotes the displacement of
the mass element of the walls at point $\mathbf r$ and time $t$,
relative to its position at rest, the displacement of the walls
can be written as the superposition of the mechanical normal modes
$\bm\xi_\alpha(\mathbf r)$ \cite{saulson}: $\mathbf u(\mathbf r,t)
= \sum \bm\xi_\alpha(\mathbf r) \, q_\alpha(t)$, with $\int
\bm\xi_\alpha(\mathbf r) \cdot \bm\xi_\beta(\mathbf r)
\rho(\mathbf r) \, dV = M\delta_{\alpha\beta}$, where $M$ and
$\rho$ are the mass and the density of the walls.

We want to study the energy transfer between two electromagnetic
normal modes of a resonator whose boundary is perturbed by an
external force when mode "1" is highly excited by an external
source at its eigenfrequency, $\hst_1 \approx A_1 \cos(\oms t)$
and $\est_1 \approx A_1 \sin(\oms t)$, with {\em constant}
amplitude $A_1$. We shall assume that only one mechanical mode
couples to the external force. Including empirical damping terms,
the equations of motion for the field amplitudes are given by:
\beq \label{eq:simsys1} \ddot{\cal H}_2 +
\frac{\omega_2}{\mathcal{Q}} \dot{\cal H}_2 + \omega_2^2 {\cal
H}_2 = -\frac{1}{2}\,\omega_2^2 q_m C^m_{21} {\cal H}_1 \eeq \beq
\label{eq:simsys2} \ddot q_m + \frac{\omega_m}{Q_m}\,\dot q_m +
\omega_m^2 q_m = \frac{f_{m}}{M} - \frac{1}{2}\frac{C^{m}_{21}}{M}
\, \hst_2\hst_1^* \eeq where $f_m(t) = \int \mathbf f(\mathbf r,t)
\cdot \bm\xi_m(\mathbf r) \, dV$, and $\mathbf f(\mathbf r,t)$ is
the volume force density which acts on the walls. The
time--independent coupling coefficient $C^{m}_{21}$ is given by
(the superscript $m$ labels the {\em mechanical} normal mode,
while the subscripts label the {\em electromagnetic} modes): \beq
\label{eq:gammap} C^{m}_{21} = \int_{{S}} (\mathbf {H}_2 \cdot
\mathbf {H}_1 - \mathbf {E}_2 \cdot \mathbf {E}_1) \, \bm \xi_{m}
\cdot d\mathbf S\,. \eeq

\begin{figure}[hbt]
 \begin{center}
 \includegraphics[scale=0.75]{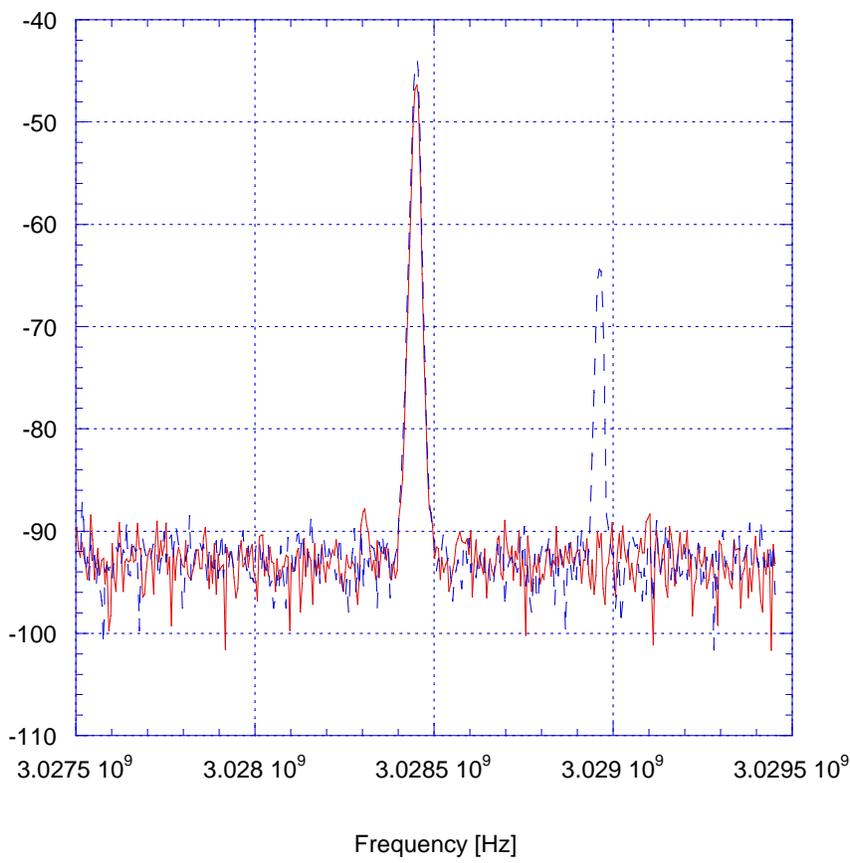}
 \caption{
 \label{fig:twopiezos}
Power trasfer between the symmetric and the antisymmetric mode
with in--phase piezo excitation (solid) and 180 degress
out--of--phase excitation (dashed)
 }
 \end{center}
 \end{figure}

The dependence of the coupling coefficient, and therefore of the
energy transfer, both on the field spatial distribution and on the
boundary deformation, has been checked using a resonator made up
of two pill--box cavities, mounted {\em end--to--end} and coupled
by a small circular aperture in their common endwall. The
perturbation of the resonator's boundary was induced by two
piezoelectric crystals mounted in the center of the two circular
endwalls. The TE$_{011}$ symmetric mode at $3$ GHz was excited by
an external rf source and the piezos were driven by a synthesized
oscillator tuned at the frequency corresponding to the
symmetric--antisymmetric mode separation ($\approx 500$ kHz). The
relative phase of the signals driving the two piezos could be set
to $0$ degrees and to $180$ degrees with an external switch. Eq.
\ref{eq:gammap} predicts for this field and for this boundary
configuration $C^{m}_{21} = 0$, for the in-phase excitation and
$C^{m}_{21} = 2$, for the excitation with $180$ degrees phase lag.
These predictions are clearly confirmed by the data shown in Fig.
\ref{fig:twopiezos}.
The explicit calculation of the coupling coefficient $C_{21}^m$
for an arbitrary deformation of the resonator volume can be done
only by numerical methods. Analytic calculations showed that for
an ideal spherical hollow resonator, excited in the fundamental
quadrupolar mechanical mode and in the TE$_{011}$ electromagnetic
mode, ${C_{21}^m} = 1.9$. More detailed calculations, made on a
realistic model of the coupled spheres, including the central
coupling cell and the e.m. input and output ports, were made by
finite element methods.

$\hst_2(t)$ is given by: \bear
\hst_2(t)=\int_{0}^{\infty} K(\tau) f_m(t-\tau)\,d\tau = \frac{1}{2\pi}\int_{-\infty}^{+\infty} \widetilde K(\omega) {\widetilde f}_m(\omega) \exp(i\omega t)\,d\omega \,;\nonumber \\
K(\tau)  = 0 \; {\mathrm{for}} \; \tau \leq 0 \,. \eear {  The
density of the external forces driving the motion of the system,
i.e. the tides generated by the passing gravitational wave, are
given by}
\begin{equation}
\label{eq:gwriem} f_i(\mathbf r,t) = -\rho(\mathbf r) R_{0i0j}(t)
x^j \,,
\end{equation}
{  where $\rho$ is the wall density, and $R_{0i0j}(t)$ are the
components of the Riemann tensor evaluated in the center of mass
of the solid. The components of the Riemann tensor can be
expressed in terms of the adimensional amplitude of the g.w.
$R_{0i0j}(t) = -1/2 \, \ddot a_{ij}(t)$.}

For a "+" polarized plane g.w travelling along the $z$ axis the
force density, in the proper reference frame attached to a
detector {  lying in the $xy$ plane}, has the form:
\begin{equation}
\label{eq:gwfor} \mathbf f(\mathbf r,t) = \frac{1}{2} \rho(\mathbf
r) \ddot a(t) \left [ -x , y, 0 \right ]\,,
\end{equation}
{  with a similar expression for the "$\times$" polarization}. The
generalized force, acting on the $m$ mechanical mode, then has the
form \bear \label{eq:gengwfor} f_m(t) = -\frac{1}{2} \,\ddot a(t)
\int_{Vol} \left[(\xi_m)_x \, x - (\xi_m)_y \, y \right]
\rho(\mathbf r) \, dV \,. \eear ${\widetilde f}_m(\omega)$ is then
given by: \beq \label{eq:fomega} {\widetilde f}_m(\omega) =
\frac{1}{2}\,M \omega^2 {\mathcal{L}} \, {\widetilde a}(\omega) \,
\eeq having defined the detector's effective length (for this
mechanical mode and wave polarization): \beq {\mathcal{L}} =
\frac{1}{M}\int_{Vol} \left[(\xi_m)_x \, x - (\xi_m)_y \, y
\right] \rho(\mathbf r)\, dV  \, \eeq The function $\widetilde
K(\omega)$ is found to be: \beq \label{eq:romega} \widetilde
K(\omega) = \frac{C^{m}_{21} \oma^2 A_1/(2M) } {
\left[\oma^2-(\omega+\oms)^2+i\,\frac{\oma}{\qst}(\omega+\oms)\right]
\left[\omega_m^2-\omega^2 + i\,\frac{\omega\omega_m}{Q_m}\right] -
\frac{(C^{m}_{21}A_1\oma)^2}{4M} }\,\exp(i\,\oms t) \, \eeq

The second term in the denominator of Eq. \ref{eq:romega} is due
to the back--action of the electromagnetic fields on the cavity
walls. Its effect is particularly important when $\oma - \oms
\approx \omega_m$. In this case optimum signal transfer does not
correspond to the maximum field amplitude $A_1$ that can be stored
in the initially excited mode (that is limited by the critical
field of the superconductor, $H_c \approx 0.2$ T for niobium).
Instead, optimum signal transfer is obtained from a lower value
that has to be optimized according to the prevailing experimental
conditions and detector characteristics. The dependence of the
signal on the energy stored in the initially excited mode will be
further discussed in Section \ref{sec:sensit} where the expected
sensitivity of the detector in various experimental conditions is
analyzed.

The average energy stored in mode number 2 is $U_2 = 1/2
|\hst_2(t)|^2$ and the power extracted from a cavity port coupled
to an external load with a coupling coefficient $\beta_2$ is $P_2
= \beta_2\,(\oma/\qst)\,U_2$.

If $a(t) = h \alpha(t)$ with $\alpha(t)$ periodic at angular
frequency $\Omega$, the expression of the output power from mode
$2$ is given by $P_2 = {\mathcal{R}}(\Omega) \, h^2$, with \beq
{\mathcal{R}}(\Omega) = \frac{1}{4} \, \frac{\beta_2
\oma}{\qst}\,\Omega^4 M^2 {\mathcal{L}}^2 |\widetilde K(\Omega)|^2
\,. \eeq

\subsection{Noise} To study how the thermal fluctuations of the
walls contribute to the output signal, we will start again from
Eqs. \ref{eq:simsys1}--\ref{eq:simsys2}, taking now the external
force $f_m(t)$ as a stochastic force with constant power spectrum
$S_{fm} = 4Mk_BT\omega_m /Q_m$ \cite{papoulis}. The output noise
power spectral density is given by: \beq S_{PP}(\omega) =
\frac{\beta_2\oma}{\qst} \, |\widetilde K(\omega) |^{2} S_{ff}
\eeq

The thermal fluctuations of the electromagnetic field itself can
be calculated in an analogus way. Starting again from Eqs.
\ref{eq:simsys1}--\ref{eq:simsys2}, we consider an external,
stochastic force coupled to the magnetic field $\hst_2(t)$, with
constant power spectrum $S_{fe}= 4kT/(\oma \qst)$. The output
noise spectral density is given by: \beq S_{\hst\hst}=
\frac{\beta_2\oma}{\qst} \, |\widetilde \chi(\omega) |^{2} S_{fe}
\eeq with $\widetilde\chi(\omega)$ given by: \beq
\label{eq:chiomega} \widetilde\chi(\omega) = \frac{\oma^2
\left(\omega_m^2-\omega^2 + i\,\frac{\omega\omega_m}{Q_m}\right)}
{
\left[\oma^2-(\omega+\oms)^2+i\,\frac{\oma}{\qst}(\omega+\oms)\right]
\left[\omega_m^2-\omega^2 + i\,\frac{\omega\omega_m}{Q_m}\right] -
\frac{(C^{m}_{21}A_1\oma)^2}{4M} } \,. \eeq We note that in the
limit of vanishing coupling, $C^{m}_{21}\approx 0$, the
contribution to the output noise of the walls vanishes, while the
contribution of the electromagnetic field fluctuations tends to be
like a simple harmonic oscillator of frequency $\oma$ and quality
factor $\qst$.

Other noise sources must also be taken into account. To operate
our device we have to feed microwave power into  mode $1$, using
an external rf source locked on mode $1$, at frequency $\oms$. The
master oscillator phase noise is filtered through the resonator
linewidth; the power spectral density has the following frequency
dependence \cite{slater}: \beq \label{eq:powdiss} S_{MO}(\omega) =
\frac{4 \beta_1 P_i/(\oms \qst)}{\left(\frac{1}{\qst}\right)^2 +
\left(\frac{\omega+\oms}{\oms} -
\frac{\oms}{\omega+\oms}\right)^2} \eeq where $P_i$ is the power
input level and $\beta_1$ is the coupling coefficient of mode $1$
to the output load. From the above equation we can estimate the
microwave power noise spectral density at the detection frequency
$\omega=\oma-\oms$: \beq \label{eq:powoma} S_{MO}(\oma-\oms)
\approx \beta_1\frac{P_i}{\oms\qst}\left( \frac{\oma}{\oma-\oms}
\right)^2 \eeq

This figure can be improved if the receiver discriminates the
spatial field distribution of the e.m. field at frequency $\oma$,
i.e. if it is sensitive only to the power excited in mode number
$2$, rejecting all contributions coming from mode number $1$. In
this way mode $1$ is decoupled from the output load and $\beta_1 =
0$. The experimental set--up, based on the use of two magic--tees
which accomplishes this issue, is discussed in detail in
\cite{rsi}. Of course, the mode discrimination cannot be ideal,
and some power leaking from mode $1$ to the detector's output will
be present. Nevertheless our previous work has demonstrated that
with careful tuning of the detection electronics we can obtain
$\beta_1 \leq 10^{-14}$ \cite{rsi}.

The input Johnson noise of the first amplifier in the detection
electronics has to be added to the previous contributions to
establish the overall noise spectral density. It can be described
by the frequency independent spectral density \cite{papoulis}:
\beq \label{eq:johnson} S_{JJ} = k_B T ( 10^{(N/10)}-1) \equiv k_B
T_{eq} \eeq where $N$ is the noise figure of the amplifier (in dB)
and $T$ the operating temperature. The equivalent temperature (or
noise temperature) $T_{eq}$, is equal to the temperature (in
Kelvin) of a 50 ohm termination at the input of an ideal noiseless
amplifier with the same gain and generating the same output noise
power.

Other noise sources (e.g. the seismic noise) are not considered in
this paper.

We shall characterize the noise in our detector by a frequency
dependent spectral density $S_h(\omega)$, with dimension Hz$^{-1}$
defined as \beq \label{eq:sh} S_h(\omega) =
\frac{S_n(\omega)}{{\mathcal{R}}(\omega)} \eeq where
$S_n=S_{PP}+S_{\hst\hst}+S_{MO}+S_{JJ}$, is the detector noise
spectral density.

\subsection{Sensitivity} Let us focus our attention on the system
based on two spherical niobium cavities working at $\oms \approx
\oma \approx 2$ GHz with a maximum stored energy in the initially
excited symmetric mode of $U_1 \approx 10$ J per cell
(corresponding to a maximum surface magnetic field $H_{max} = 0.1$
T, half the critical field of niobium). This is  a small--scale
system with an effective length of 0.1 m and a typical weight of 5
kg. The lowest quadrupolar mechanical mode is at $\omega_m \approx
4$ kHz. In the following, we shall consider an equivalent
temperature of the detection electronics $T_{eq} = 1$ K.

\begin{figure}[hbt]
 \begin{center}
 \includegraphics[scale=0.75]{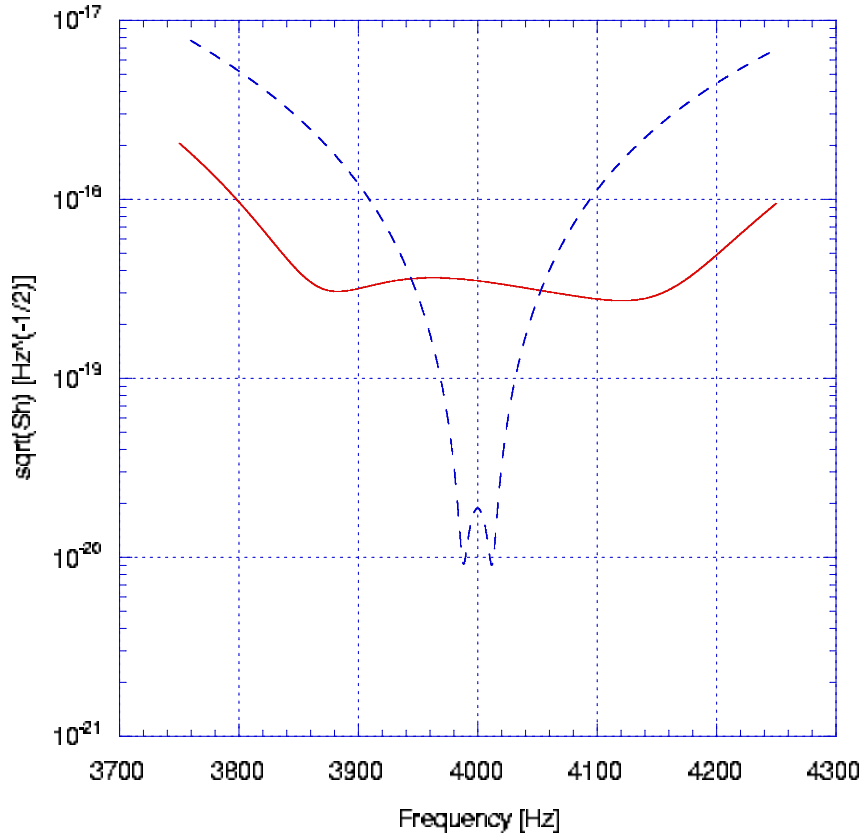}
 \caption{
 \label{fig:hmin4}
 Calculated system sensitivity ($\omega_m \approx \oma - \oms \approx 4$ kHz, ${\mathcal{Q}}=10^{10}$, $T=1.8$ K,
 $T_{eq}=1$ K, and a) $Q_m=10^{3}$, stored energy $U\approx 10$ J per cavity (solid line);
 b) $Q_m=10^{6}$, stored energy $U\approx 0.1$ J per cavity (dashed line))
 }
 \end{center}
 \end{figure}

A possible design of the detector uses both the mechanical
resonance of the structure, and the e.m. resonance. Due to the
tuning system, the detection frequency can be made equal to the
mechanical mode frequency $\omega_m \approx \oma - \oms$.  The
expected sensitivity of the detector for $\oma - \oms = \omega_m =
4$ kHz is shown in figure \ref{fig:hmin4}, for a mechanical
quality factor $Q_m =10^3$ (solid line) and $Q_m=10^6$ (dashed
line). Note that, in the two cases, the optimum sensitivity is
obtained with different values of stored energy. In both cases the
stored energy has been optimized for maximum detector bandwidth.
When the mechanical quality factor is higher ($Q_m=10^6$) the
stored energy has to be maintained much under the maximum allowed
value.

When $\oma - \oms = \omega_m$, the detector sensitivity is limited
by the walls thermal motion. In this case, a lower $T_{eq}$ would
increase the detection bandwidth.

\begin{figure}[hbt]
 \begin{center}
 \includegraphics[scale=0.75]{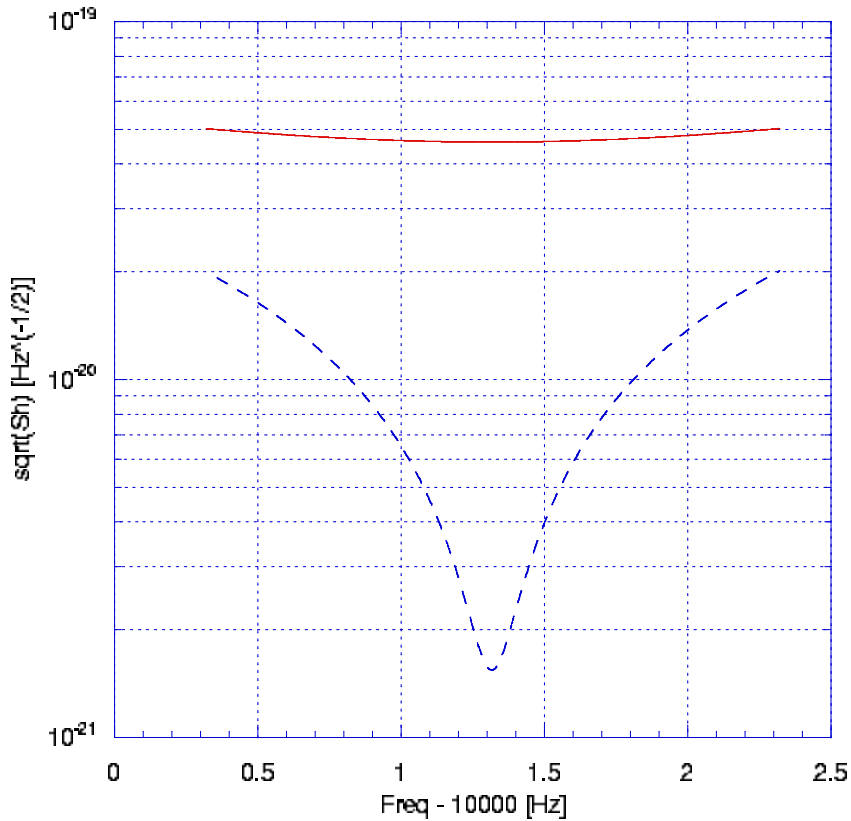}
 \caption{
 \label{fig:hmin10}
 Calculated system sensitivity ($\omega_m \approx 4$ kHz, $\oma - \oms \approx 10$ kHz, ${\mathcal{Q}}=10^{10}$,
 $T=1.8$ K, $T_{eq}=1$ K, stored energy $U\approx 10$ J per cavity and a) $Q_m=10^{3}$ (solid line);
 b) $Q_m=10^{6}$ (dashed line))
 }
 \end{center}
 \end{figure}

Since our detector is based on a double resonant system (the
mechanical resonator and the electromagnetic resonator) it can be
operated also for frequencies $\oma-\oms \neq \omega_m$. At
frequencies $\oma-\oms \leq 1$ kHz the master oscillator phase
noise will, in general, completely spoil the system sensitivity,
while at frequencies $\oma-\oms \geq 10$ kHz the noise coming from
the detection electronics will dominate. The expected sensitivity
of the small--scale detector for $\oma - \oms = 10$ kHz is shown
in figure \ref{fig:hmin10}.

\begin{figure}[hbt]
 \begin{center}
 \includegraphics[scale=0.75]{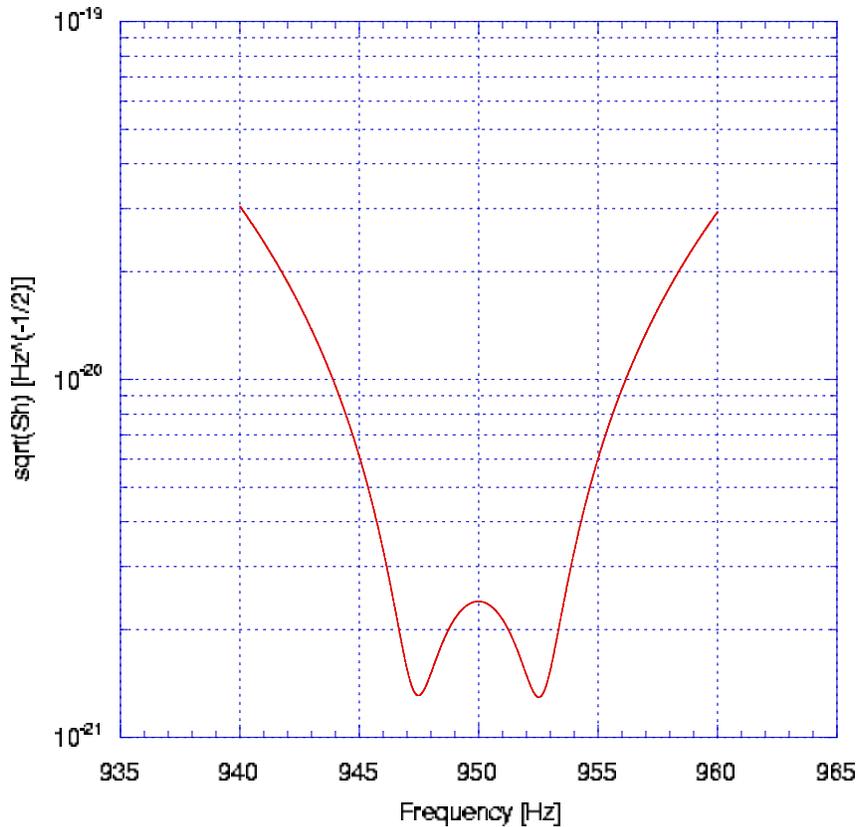}
 \caption{
 \label{fig:hmin1}
 Calculated system sensitivity ($\omega_m \approx \oma - \oms \approx 1$ kHz, ${\mathcal{Q}}=10^{10}$, $Q_m=10^{6}$, $T=1.8$ K, $T_{eq}=1$ K, stored energy $U\approx 1$ J per cavity)
 }
 \end{center}
 \end{figure}

In order to increase the expected sensitivity a large--scale
system has to be developed. A possible design could be based on
two spherical cavities working at $\oms \approx \oma \approx 500$
MHz, with $\omega_m \approx 1$ kHz. This system could have a
maximum stored energy of $U_1 \approx 1200$ J per cell, an
effective length of 0.4 m and a typical weight of 2300 kg. With a
reasonable choice of system parameters one could obtain the
sensitivity shown in figure \ref{fig:hmin1}, for the
double-resonance case ($\oma-\oms=\omega_m$). As in the previous
(small--scale) case the energy store in the initially excite mode
has been optimized for maximum bandwidth, and it has to be much
less then the maximum allowed. Also in this case lowering $T_{eq}$
corresponds to an increase of the detection bandwidth.

\begin{figure}[hbt]
 \begin{center}
 \includegraphics[scale=0.75]{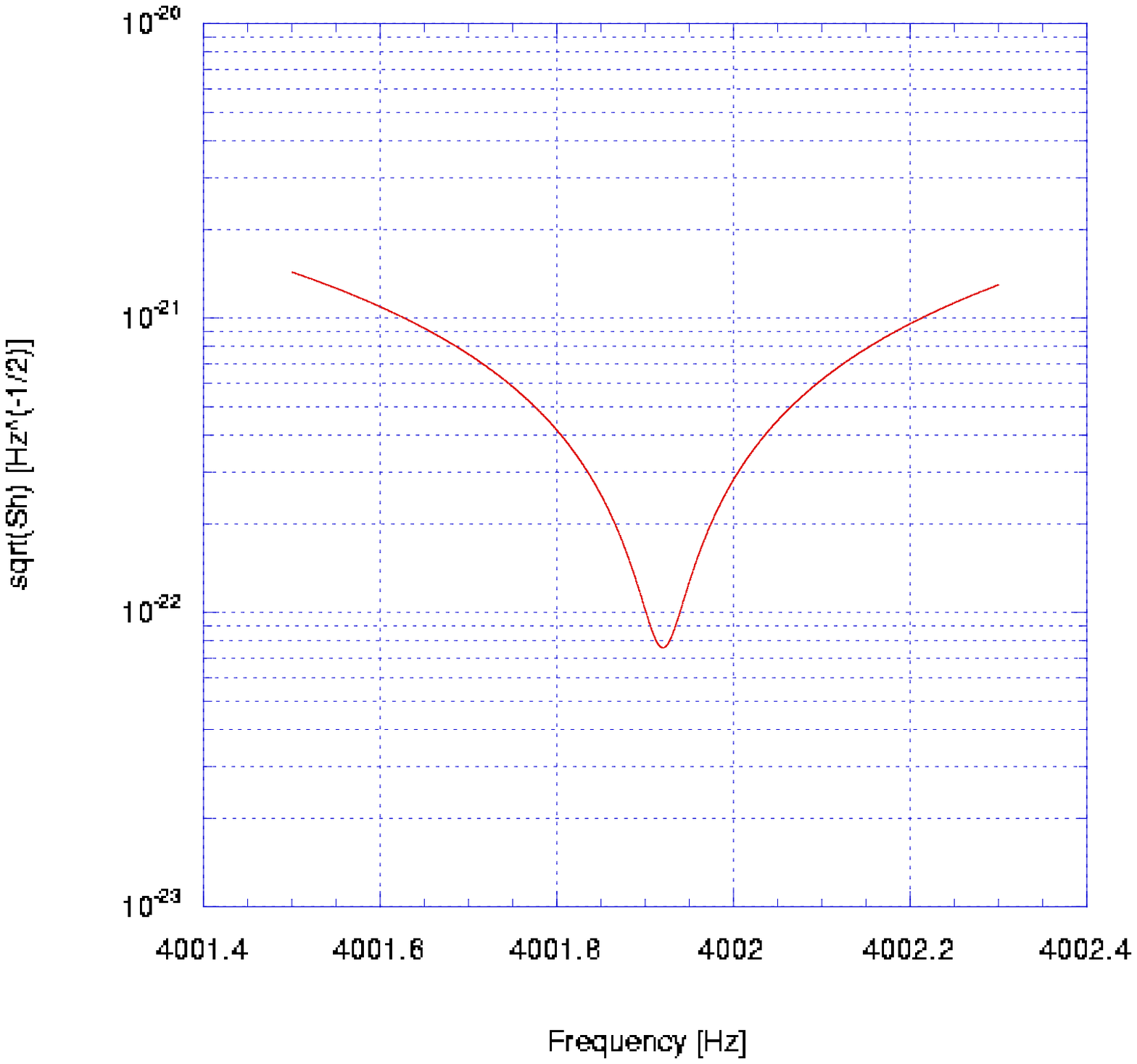}
 \caption{
 \label{fig:hmin10g}
 Calculated system sensitivity ($\omega_m \approx 1$ kHz, $\oma - \oms \approx 4$ kHz, ${\mathcal{Q}}=10^{10}$,
 $Q_m=10^{6}$, $T=1.8$ K, $T_{eq}=1$ K, stored energy $U\approx 1200$ J per cavity)
 }
 \end{center}
 \end{figure}

Obviously the large--scale system could also be used at higher
frequencies; in this case a good sensitivity can be achieved in a
narrow detection bandwidth (see Fig. \ref{fig:hmin10g}).

It is worth noting that the narrow detection bandwidth is not an
unavoidable drawback of the system. Actually its value is
determined by the coupling coefficient of the antisymmetric mode
at the $\Delta$ port of the output magic--tee, and can be adjusted
changing this coupling. This corresponds to changing (lowering)
the quality factor of the antisymmetric mode, leaving the quality
factor of the symmetric mode unaffected. Of course an increased
bandwidth corresponds to a lower sensitivity, since the latter is
proportional to the antisymmetric mode quality factor. The
possibility to increase the detection bandwidth is also
interesting for other possible applications of this detection
technique which is based on the parametric frequency conversion
between two electromagnetic modes in a cavity (for example,  in
connection with recently proposed detectors based on the dual
resonator concept \cite{dual1, dual2, dual3}).

\section{Data Analysis}
Several aspects of MAGO data analysis need to be investigated.

The data-analysis  for both single and multiple MAGOs will be akin
of those used for acoustic detectors \cite{Finn}, in view of the
narrowband MAGO response.

Data analysis strategies for MAGOs operating above $f \sim 4\cdot
10^3$ Hz will be strictly {\em non parametric}, in view of the
present lack of GW source/signal models in this frequency range.

Optimum MAGO tuning schedules for detecting unknown
(sub)millisecond pulsars \cite{ms_PSR}, \cite{ms_PSR1} will be
obviously a most relevant issue to be investigated.

A "xylophone" of MAGOs tuned in the range between $10^3$ Hz and
$10^4$ Hz might be able to {\em both} detect {\em and estimate}
the chirp-mass \cite{Finn1} of galactic light\footnote{
%+++++++++++
The highest GW frequency radiated during binary inspiral is twice
the orbital frequency of the last stable circular orbit, and is
\cite{Finn1} $f \sim 4\cdot 10^3 (M_\odot/M)$ [Hz] . }
%+++++++++++
BH-MACHO binaries \cite{MACHO}, which would produce a distinct
signature in the xylophone output \cite{WG}. The expected
(optimistic) event-rate might be non negligible \cite{Finn}.

Once MAGO prototypes will be up and running, a primary task will
be to investigate the experimental noise PSD and transfer
function, by comparison to their theoretical counterparts.

\section{Future Developments}
The sensitivity of the device currently under test, consisting of
a couple of nearly spherical niobium cells, 2~mm thick, whose
total mass is approximately 4.5~kg, might be improved in several
ways. Among these the most promising are:
\begin{itemize}
    \item an increase of the mechanical quality factor of the structure;
    \item an increase of the electromagnetic quality factor of the superconducting cavities;
    \item a lower noise temperature of the detection electronics;
    \item the increase of the size (and mass) of the detector;
    \item the use of several detectors in an {\em array} or in a {\em xylophone}.
\end{itemize}
All these items are discussed in the following.

\subsection{Increasing the mechanical quality factor} \label{sec:mechq}
Although the $S/N$ ratio does depend on several parameters, we can
state that if the noise power spectral density is limited by
thermal noise, the minimum $S_h$ is proportional to the square
root of the mechanical quality factor, $(Q_m)^{1/2}$.

The mechanical quality factor depends both on the \emph{intrinsic}
mechanical energy dissipated in the material of the resonator and
on the energy dissipated through the coupling to the external
environment. The intrinsic quality factor of niobium at low
temperature and few kHz is $Q_m \sim 10^8$. Since our detector is
immersed in a bath of subcooled superfluid helium, improvements in
$Q_m$ can be expected thanks to carefully designed liquid helium
bath and suspension system\footnote{In this document we consider
this cryogenic scheme. Different schemes, not based on the use of
a liquid helium bath (e.g. pulse tube refrigerator), might be
foreseen.}.

The limit imposed by mechanical dissipation in the superfluid
helium bath is $Q_m \sim 10^6$ for the present small--scale
detector (see sec. \ref{sec:absorption}).

This figure can be further improved for a larger detector ($M
\agt 10^4$~kg). In this case, the expected
liquid--helium--limited mechanical quality factor should be $Q_m
\sim 10^7$ (see sec. \ref{sec:absorption}).

We remark that the above limits are essentially due to the will of
using a well established refrigeration scheme for the
superconducting cavity. In principle different schemes could be
developed which do not make use of a bath of coolant in contact
with the cavity (e.g. pulse tube refrigerators). Obviously this
requires a consistent R\&D effort which is not, for the time
being, in the line of our proposal which is entirely based on
already existing technologies.

\subsection{Increasing the electrical quality factor} \label{sec:eleq} The
gain in sensitivity due to the electrical quality factor, $\qst$,
is less trivial to discuss. First of all a distinction should be
made between the electromagnetic quality factor of the
\emph{symmetric} (pump) mode and of the \emph{antisymmetric}
(detection) mode\footnote{We remind that our device is based on
the principle of parametric conversion of power between the two
electromagnetic modes of the coupled resonators: the symmetric
mode at frequency $\oms$ and the antisymmetric mode at frequency
$\oma$, induced by an external perturbation of the cavity boundary
at frequency $\Omega \sim |\oma - \oms|$}. The quality factor of
the pump mode, $\qst_1$, strongly affects the carrier (oscillator)
phase noise and the energy stored in the cavity for a given input
power. The quality factor of the detection mode, $\qst_2$, affects
the shot noise power spectral density and the detection bandwidth.

For the sake of simplicity in our proposal these two quality
factor were taken as equal: $\qst_1=\qst_2=\qst$. This is
certainly a reasonable assumption as far as the \emph{intrinsic}
quality factor of the niobium spheres: $\qst_0 = G/R_s$ is
considered. In fact $\qst_0$ depends only on the geometrical shape
of the resonator, on the electromagnetic resonant mode and on the
surface quality. The overall
(\emph{loaded}) quality factor of a resonator, though, is given by
$\qst = \qst_0/(1+\beta)$, where $\beta$ is the coupling
coefficient to the external load\footnote{The coupling coefficient
$\beta$ determines the impedance match between the input (for the
pump mode) or output (for the detection mode) impedance of the
cavity $Z_c(\omega)$, and the impedance of the external load
$Z_0$. The ratio $Z_c/Z_0$ is simply the coupling coefficient
$\beta$.}.

As a consequence, it is in general convenient to keep the quality
factor of the pump mode $\qst_1$ as high as possible, in order to
have the minimum power dissipation for a given energy stored in
the detector (or the maximum stored energy for a given level of
power dissipation) and the minimum carrier phase noise at the
detection frequency. This requirement is fulfilled when $\beta_1
\sim 1$ (critical coupling).

On the other hand, the value of the detection mode quality factor,
$\qst_2$, is a parameter that has to be set to the optimum value
(which depends on the specific experimental situation) by tuning
the output coupling coefficient $\beta_2$. To be more specific,
since shot noise and amplifier back--action noise are roughly
proportional to $(\qst_2)^{-1/2}$, when either of these is the
dominant noise contribution an higher $\qst_2$ gives better
sensitivity (and, in general, a small detection bandwidth). On the
contrary, when sensitivity is limited by other noise sources (e.g.
thermal noise) it is convenient to lower $\qst_2$ (increasing
$\beta_2$) to open the bandwidth.

Improvements in the intrinsic quality factor, $\qst_0$, can be
expected thanks to specific Nb surface treatments that give a
lower $R_s$. Quality factor values $\qst_0\agt 10^{11}$ have
already been measured on several prototypes of elliptical
accelerating cavities ($R_s \sim 1$~nOhm) both bulk niobium
\cite{safa}, and sputtered niobium on copper \cite{sergio}. Since
those cavities were run on the TM$_{010}$ accelerating mode, we
can expect $\qst_0 \sim 10^{12}$ (on the TE$_{011}$ mode) with a
comparable surface quality, due to the increased geometrical
factor\footnote{The ratio between the geometric factor of a
spherical cavity operating in the TE$_{011}$ mode and an
elliptical cavity operating in the TM$_{010}$ mode is
$G_{TE}/G_{TM} \sim 3.4$}.

\subsection{Increasing the performance of the detection electronics}
\label{sec:elect} A further headway can be foreseen using a low
noise amplifier. The noise temperature of a cryogenic,
commercially available, high--frequency amplifier (GaAs or InP HEMT device)
is $T_{eq} = 2$~K.
Bearing in mind that we want to amplify an e.m. field at a
frequency around $\omega/(2\pi) \sim 2$~GHz (much greater than the
detection (g.w.) frequency $\Omega_{gw}=\oma-\oms$) one can easily
see that the minimum (quantum limited) noise temperature of an
amplifier is $T_{eq}=\hbar\omega/k_B\sim 0.1$~K.

This apparently high noise temperature though is not a drawback,
since the parametric transducer operates the up--frequency
conversion from low energy quanta $\hbar\Omega_{gw} \rightarrow
\hbar\omega$ \emph{with no added noise}. In this process the gain
(in energy) is $\omega/\Omega_{gw}\sim 10^6$ \cite[sec.
7.9--7.11]{louisell2}.

In other words, the noise temperature \emph{seen at the input of
the parametric up--converter} is $T_{eq}\sim 0.1\times
\Omega_{gw}/\omega\sim 10^{-7}$~K.

\subsection{Increasing the size of the detector} \label{sec:MAGO}
Starting from the experience gathered with the existing
small--scale prototype, a plausible detector (inner radius of the
spherical cell $a=0.2$~m, thickness $w=5$~mm) could be designed
and built\footnote{There are no obstacles, in principle, to
envisage an even larger detector, facing, obviously, major
technical challenges. Major differences lay in the greater mass,
greater energy stored, lower detection frequencies and possibly
fabrication methods (e.g. sputtered Nb on a substrate).}.

Such detector might have the following parameters: overall mass $M
\sim 44$~kg (approx. a ten--fold increase with respect to the
present small--scale prototype), characteristic length
$\mathcal{L} \sim 1$~m, working rf frequencies $\omega/(2\pi) \sim
1$~GHz, stored energy $U \sim 210$~J, detection frequency range
$2$~kHz $\alt \Omega_{gw}/(2\pi) \alt 10$~kHz.

Being the detector tunable in frequency, we shall divide its use
into two different working ranges: a \emph{Near Mechanical
Resonance} range (hereafter NMR) and an \emph{Off Mechanical
Resonance range} (OMR). NMR operation can only be used next to
both quadrupolar mechanical resonances of the spherical shell (at
frequencies $f_1 \sim 2$~kHz and $f_2 \sim 8$~kHz). The two ranges
require quite different parameter sets and lead to optimized use
of the detector, depending on the sources under investigation.

\begin{figure}[hbt]
 \begin{center}
\includegraphics[scale=0.75]{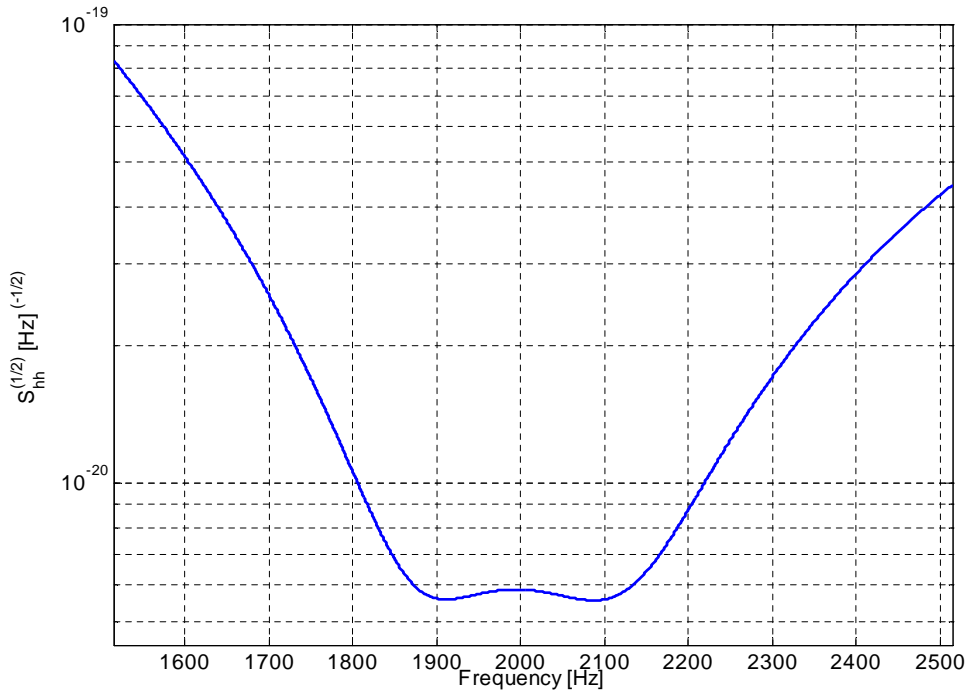}
 \caption{Near mechanical resonance PSD of one MAGO. Minimum
 $(S_h)^{1/2} \sim 6\times 10^{-21}$~(Hz)$^{-1/2}$, bandwidth $\Delta f \sim 350$~Hz.
 Mechanical quality factor $Q_m=10^6$ (limited by losses in liquid helium),
 electrical intrinsic quality factor $\qst_0 = 8.5\times10^{11}$ ($R_s = 1$~nohm),
 amplifier equivalent temperature $T_{eq}=0.05$~K (quantum limit).
 \label{fig:NMRMAGO}
 }
 \end{center}
 \end{figure}

\begin{figure}[hbt]
 \begin{center}
\includegraphics[scale=0.75]{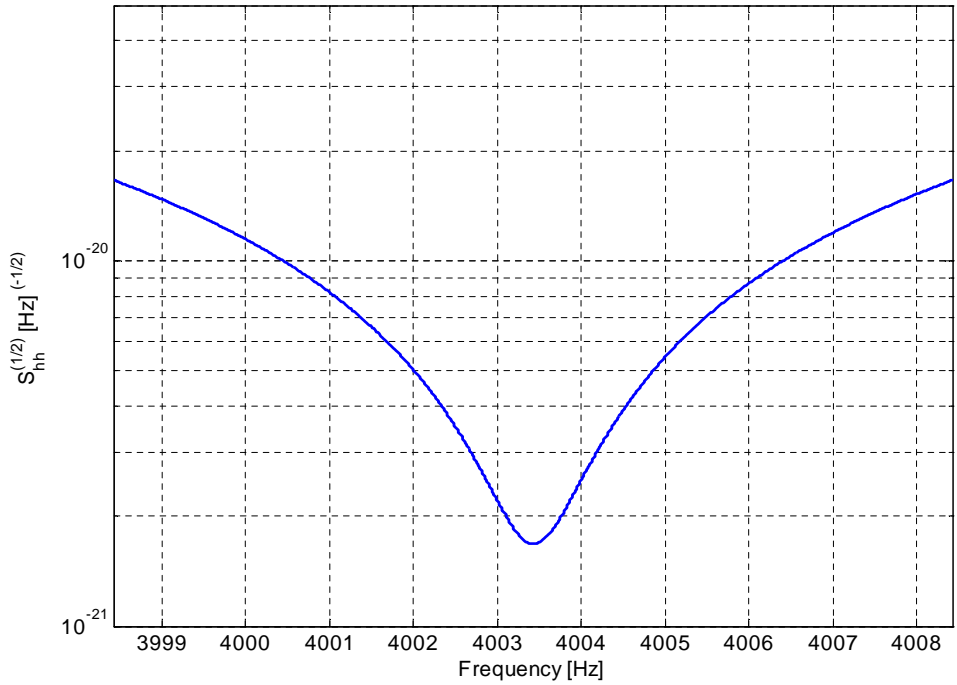}
 \caption{Off mechanical resonance PSD of one MAGO. Minimum
 $(S_h)^{1/2} \sim 1.6\times 10^{-21}$~(Hz)$^{-1/2}$, bandwidth $\Delta f \sim 1$~Hz.
 Other parameters as in fig. \ref{fig:NMRMAGO}.
 \label{fig:OMRMAGO}
 }
 \end{center}
 \end{figure}

Fig. \ref{fig:NMRMAGO} and fig. \ref{fig:OMRMAGO} show system
sensitivity calculated with $Q_m=10^6$ (limited by mechanical
dissipation in liquid helium), $\qst_0=8.5\times10^{11}$ and a
nearly quantum limited electronics $T_{eq}=0.05$~K, in NMR and OMR
mode. All other detector parameters are optimized for NMR and OMR
operation. As expected, the use of $T_{eq}=0.05$~K for the noise
temperature of the amplifier does not bring a dramatic increase in
the peak performance of the system, whereas the bandwidth is
strongly enhanced.

We remind that the bandwidth of the detector can be adjusted also
by means of the electromagnetic coupling ($\beta_2$) as discussed
in the previous section. Adjusting the electromagnetic coupling is
of paramount importance because it allows trading maximum
sensitivity versus bandwidth and is particularly useful when the
maximum sensitivity is limited by thermal noise (NMR operation).

\subsection{Increasing the number of the detectors} \label{sec:array} A
possibly nicer extension for the detector discussed in the
previous section is given by multi--detector operation to achieve
low false--alarm probabilities in coincidence operation.
Furthermore, since MAGO can be designed to work at \emph{any}
chosen frequency in the range $10^3$--$10^4$~Hz \emph{with the
same sensitivity}, a "xylophone" of MAGOs might be able to both
detect and estimate the chirp mass of galactic light BH--MACHO
binaries, which would produce a distinct signature in the
xylophone output.

The expected (optimistic) event--rate might be not negligible (see
section \ref{sec:MACHO}).

Several authors have already pointed out the usefulness of array
detectors, both for source localization and for improving the
overall signal information (multimode transducer) \cite{frasca94,
frasca96, frasca97}.

The SNR of $N$ optimally oriented detectors, sharing the same
sensitivity band is approx. $N$ times the SNR of the single
detector \cite{frasca94}. A set of 10 MAGOs would therefore have a
peak $(S_h)^{1/2} \sim 10^{-22}$~(Hz)$^{-1/2}$.

\begin{figure}[hbt]
\begin{center}
\includegraphics[scale=0.75]{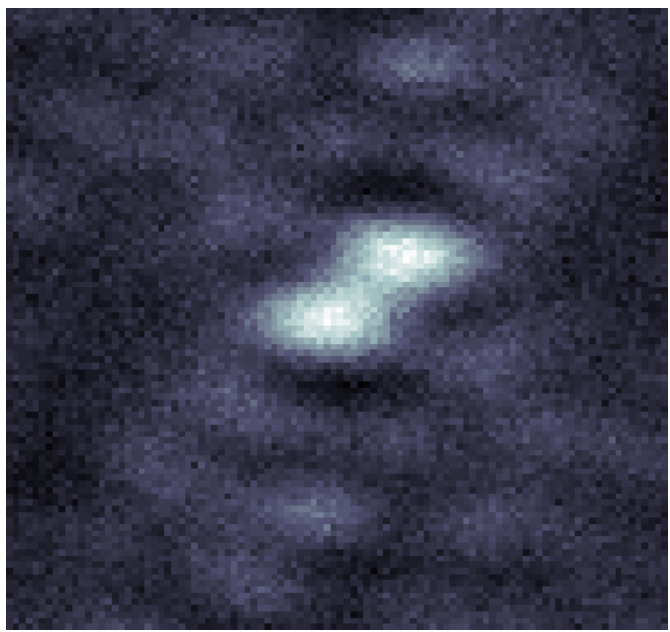}
\caption{Reconstructed image (simulated) from a network of 10 MAGO
detectors, with $SNR=0.5$ for the single detector. The angular
distance between the two sources is $\Delta\theta = \Delta\phi =
0.1\degree$ and the shown portion of the sky is $1\degree$ square.
\label{fig:image} }
\end{center}
\end{figure}

In building a MAGO coincidence network, several features of the
small single detector add to the many known benefits of arrays:

\begin{itemize}
    \item[a)] The expected cost and requirements confer a non negligible freedom
    in choosing the installation sites and financing;
    \item[b)] There is a greater freedom in the choice of the material;
    \item[c)] Short thermal cycle allow faster development phase and more flexible management of the instrument;
    \item[d)] High frequency operation is preferable from an instrumental point of view
    because many problems like vibration isolation, transducer optimization and amplifier
    matching are easier to solve;
    \item[e)] MAGO size makes it possible to more easily achieve optimal orientation
    among distant detectors. It makes also possible to host a local array in a reasonably
    limited area. This feature is particularly important in view of the detection of
    a stochastic background of GWs (see sec. \ref{sec:stoch}).
    \item[f)] Thanks to the frequency and bandwidth tuning ability, one could have
    a nearly complete overlap of the sensitivity, allowing narrow-band/high-sensitivity
    detection mode (OMR operation) not be a strong limitation. This feature might
    play a role in the detection of GW signals from binary mergers (see sec. \ref{sec:mergers}).
    \item[g)] High frequency signal sensitivity is apt to constructing high resolution
    images (see fig. \ref{fig:image}).
\end{itemize}

\end{document}